%% file: main.tex
\documentclass[aps,prl,twocolumn,superscriptaddress]{revtex4-2}

\usepackage{breqn}
\usepackage{graphicx}
\usepackage{dcolumn}   
\usepackage{bm}        
\usepackage{amssymb}   
\usepackage{soul}
\usepackage{gensymb}
\usepackage{enumerate}
\usepackage{multirow}
\usepackage{siunitx}
\usepackage{makecell}
\usepackage{pifont}

\usepackage[dvipsnames]{xcolor} 
\usepackage[breaklinks=true,colorlinks=true,linkcolor=blue,urlcolor=blue,citecolor=blue]{hyperref}
\def  \rpol {\rho_\mathrm{pol}}
\def  \nsep {n_\mathrm{sep}}
\def  \grillix {\texttt{GRILLIX}{}}
\usepackage{float}
\begin{document}

\title{Turbulent Nature of the Quasicontinuous Exhaust Regime for Fusion Plasmas}

\author{Kaiyu Zhang}
\email{kaiyu.zhang@ipp.mpg.de}
\author{Wladimir Zholobenko}
\author{Andreas Stegmeir}
\author{Michael Faitsch}
\author{Konrad Eder}
\author{Christoph Pitzal}
\author{Frank Jenko}
\author{ASDEX Upgrade Team}
\thanks{See author list of T. Pütterich et al, 2026 Nucl. Fusion 66 116002 https://doi.org/10.1088/1741-4326/ae61c8}
\affiliation{Max Planck Institute for Plasma Physics$,$ Boltzmannstr.~2$,$ 85748~Garching$,$ Germany}

\begin{abstract}

{We demonstrate a turbulence mechanism that reconciles high plasma confinement with efficient heat exhaust --- a central challenge for fusion energy. Global two-fluid turbulence simulations of the reactor-relevant Quasicontinuous Exhaust regime on the ASDEX Upgrade tokamak reveal that a quasicoherent mode drives mesoscopic oscillations of the pedestal boundary across the magnetic separatrix and ejects ballistic filaments (blobs), reproducing both the mean profiles and turbulent fluctuations observed experimentally. 
This behavior arises from a synergistic interplay between kinetic ballooning modes and resistive X-point modes straddling the separatrix. 
These first-principles results place extrapolations to future fusion reactors on a firm physical footing.} 

\end{abstract}

\maketitle

Prospective magnetic confinement fusion reactors must reconcile high energy confinement with tolerable heat exhaust to achieve net gain while protecting plasma-facing components \cite{siccinio2020demo}. 
After years of exploration \cite{Viezzer2018}, experiments across multiple tokamaks \cite{harrerParameterDependencesSmall2018,labit2019dependence,faitsch2025quasi} have identified an operational “sweet spot” called the Quasicontinuous Exhaust (QCE) regime, accessed \cite{dunne2024quasi,faitsch2023analysis} through strong plasma shaping and high plasma density $\nsep$ at the magnetic separatrix. 
QCE integrates H-mode \cite{wagner1982regime} operation with reactor-relevant heat exhaust solutions: Type-I edge-localized modes (ELMs) \cite{zohm1996edge} are naturally suppressed \cite{harrerParameterDependencesSmall2018}, the scrape-off layer (SOL) heat flux decay length \cite{Eich2013} is broadened \cite{faitsch2021broadening}, and the high $\nsep$ favors detached operation \cite{wischmeier2015high}.
These merits are phenomenologically attributed to: (i) a quasicoherent mode (QCM) across the separatrix \cite{mazurenko2002experimental,kalis2023experimental,terry2005transport}, whose sufficient transport tailors the pedestal and prevents ELMs, and (ii) intense filamentary blobs \cite{d2011convective} in the SOL, whose propagation broadens $\lambda_q$ \cite{faitsch2021broadening}.
The extrapolation to future devices is hindered by the hitherto unresolved nature of the QCM and blobs, i.e.~the generation, interaction, and regulation of those dynamics.
A first-principles understanding requires resolving turbulence across the separatrix, where nonlinear, electromagnetic, and intermittent dynamics are further compounded by {the transition in magnetic topology.} 

\begin{figure}[b] 
\centering
    \centering
    \includegraphics[trim=0.2cm 0.6cm 0.2cm 0.1cm, clip, width=0.982\linewidth]{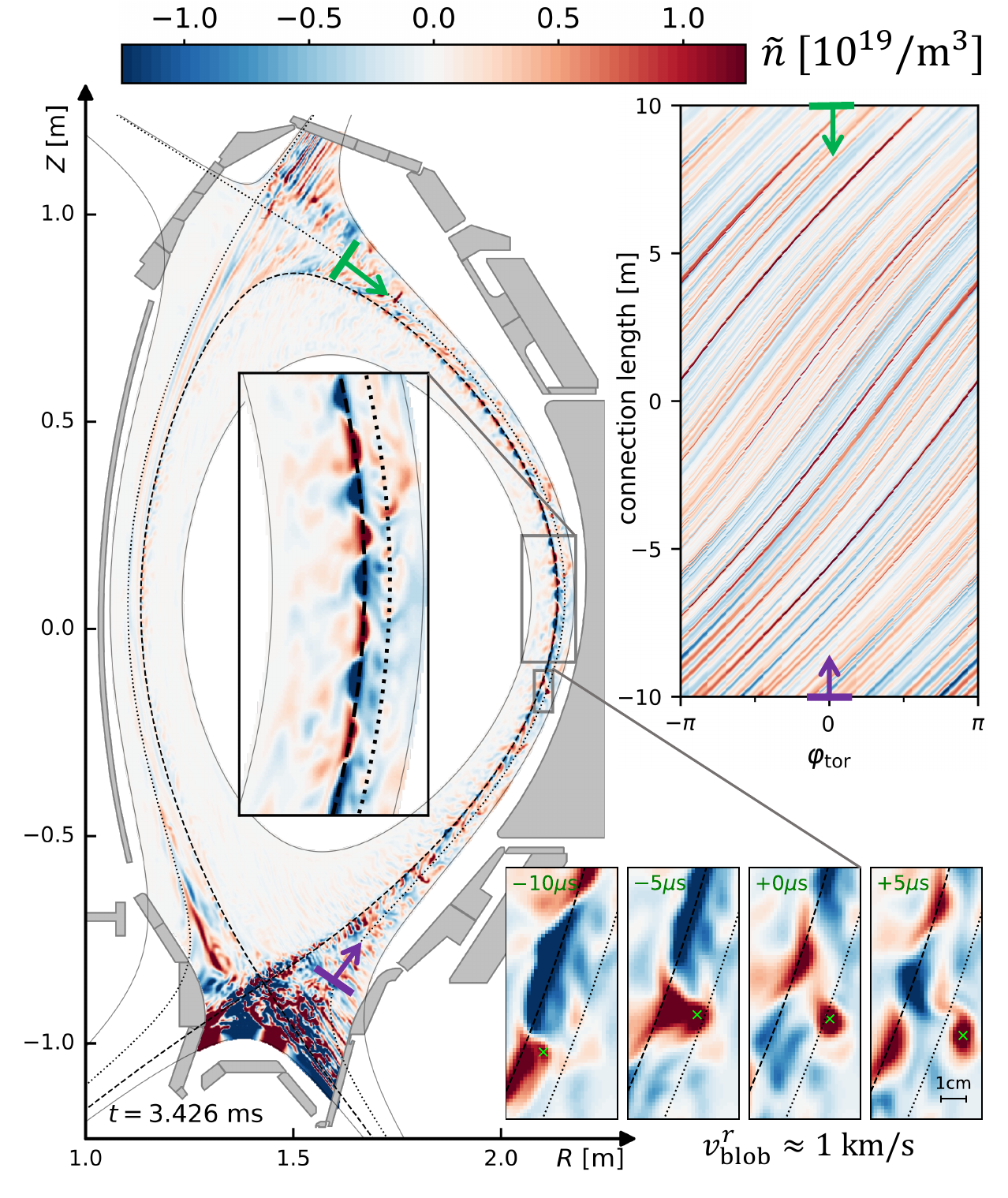}
    \caption{Composite view of {modeled} plasma density fluctuations. 
    The color scale is saturated near the divertors.
    The top-right inset shows toroidal fluctuations; the bottom-right inset shows the time series of the QCM ejecting blobs.}
    \label{fig:snaps}
\end{figure}

Here we address the challenge with the 
two-fluid electromagnetic turbulence code \texttt{GRILLIX} \cite{stegmeir2019global,zhang2024magnetic,zhang2025dynamical,zholobenko2024tokamak,zholobenkoFast}, well suited for the highly collisional QCE regime. 
The code adopts a locally field-aligned method \cite{stegmeirFieldLineMap2016, stegmeirAdvancesFluxcoordinateIndependent2017} to maintain turbulence-relevant resolution across the separatrix and X-point \cite{stegmeirAnalysisLocallyalignedNonaligned2023}.
{The background and fluctuating components of plasma quantities evolve together, and are fully incorporated into the evolution equations and physical coefficients.}
Our simulation employs the magnetic equilibrium of ASDEX Upgrade (AUG) discharge \#36165 at {$6.1\si{s}$} in the QCE phase \cite{faitsch2021broadening} reproducing the experimentally observed QCM and SOL blobs.
Fig.~\ref{fig:snaps} illustrates this with density fluctuations $\tilde{n}=n-\bar{n}$, {where the overbar denotes a toroidal and time average over the saturated phase, \mbox{$3.4$--$3.9~\si{ms}$}, used consistently throughout this study.}
The left panel shows a poloidal cross-section of the simulation domain, which includes the separatrix (dashed) and a secondary separatrix (dotted), yielding four parallel boundary regions treated with volume penalization \cite{stegmeir2019global, stegmeirAnalysisLocallyalignedNonaligned2023}.
The QCM is localized near the separatrix on the low-field side, displaying sinusoidal fluctuations and a coherent mode structure.
At the outboard mid-plane (OMP), the poloidal wavenumber is $k_{\mathrm{pol}}\rho_{s}=0.033$, within the typical range of QCM measured in AUG \cite{kalis2023experimental}, where $\rho_{s}$ is the sound Larmor radius.  
In the SOL, the outer layer of the QCM evolves into blobs, which propagate ballistically into the far-SOL with radial velocity around $1 \, \si{km/s}$, in good agreement with QCE experiments \cite{grienerContinuousObservationFilaments2020}.
These blobs have a perpendicular scale of $1 \, \si{cm}$ and are strongly elongated with parallel extents over $10 \, \si{m}$ along the magnetic field. 
This is illustrated by the top-right inset, which shows the fluctuations at the secondary separatrix versus toroidal angle $\varphi_{\mathrm{tor}}$, with the poloidal coordinate mapped to the connection length to the OMP. 
Such filamentary structures extend from the OMP to the divertor plates.

\begin{figure} 
    \centering
    \includegraphics[trim=0.2cm 0.2cm 0.2cm 0.2cm, clip, width=1\linewidth]{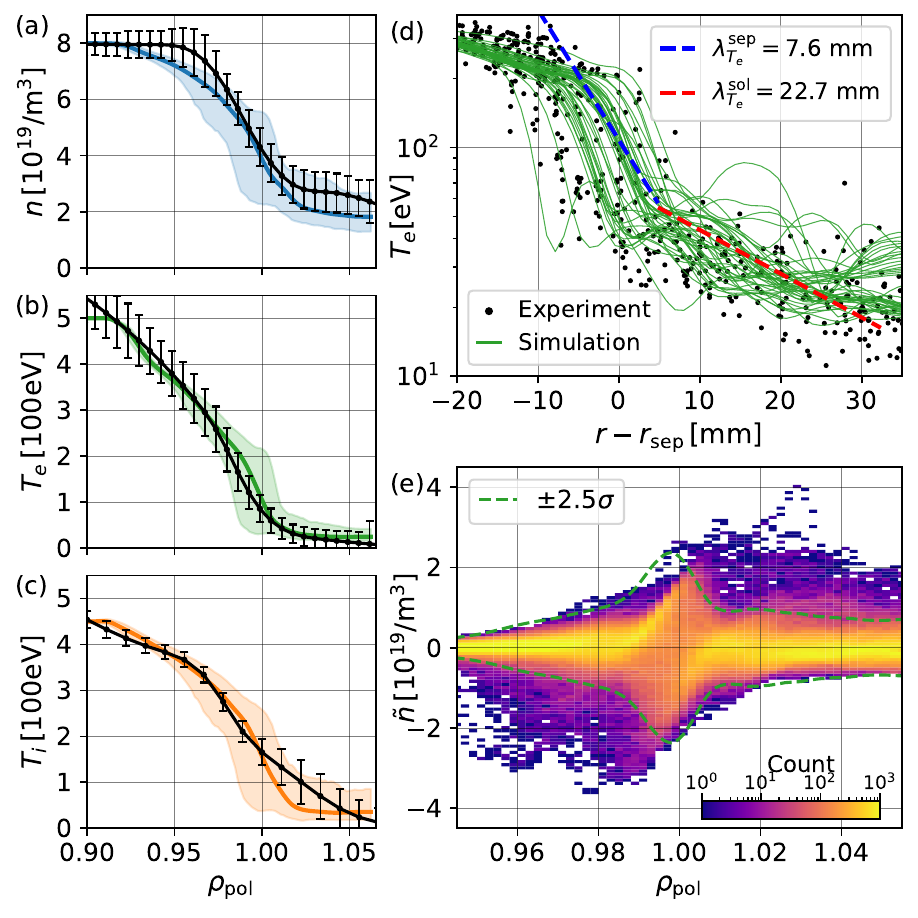}
    \caption{{OMP profiles.}
    (a-c) $n$, $T_e$, and $T_i$ vs. experimental estimates (black markers, error bars indicate uncertainties). Shaded bands show the fluctuation envelope spanning the \(1\%\)--\(99\%\) quantiles.
    (d) Fluctuating $T_e$ sampled every $0.01\, \si{ms}$ over $0.5\, \si{ms}$ on a logarithmic y-axis, compared to Thomson scattering. The blue and red dashed lines denote linear regressions of the mean $T_e$ in the simulation.
    (e) Histogram of $\tilde{n}$. The dashed lines mark the $\pm 2.5\sigma$ thresholds.
    }
    \label{fig:profiles}
\end{figure}

The plasma profiles evolve according to these dynamics and, in the saturated state, agree well with the experiment.
Fig.~\ref{fig:profiles}(a--c) compares the mean density and temperature profiles with experimental integrated data analysis (IDA) \cite{fischer2010integrated} at OMP. 
{Adaptive numerical sources are applied at $\rpol<0.92$ to match the experimental core conditions,
whereas a zero-Neumann boundary condition is used at $\rpol=1.06$, allowing the profiles to evolve.}
{
The core boundary particle source is negligible in steady state, while the heat source saturates at $3.2 \si{MW}$.}
Ionization/recombination sources are obtained by coupling to a three-moment neutral gas model with recycling boundary conditions \cite{eder2025self}, which results in a high $\nsep=4\times10^{19}/\si{m^3}$ free of ad hoc manipulation.
For parallel viscosity and heat conduction, we apply neoclassical corrections \cite{Hirshman1981,Rozhansky2009,zholobenko2024tokamak} and a Landau-fluid closure \cite{hammett1990fluid,pitzal2023landau}, respectively, extending the Braginskii model \cite{braginskii1965transport} to a wider range of pedestal collisionalites, while eliminating any free parameters from the model that could otherwise be tuned to adjust the profiles.
The first principles profiles of $n$, $T_e$, and $T_i$ remain within experimental uncertainty across the edge and SOL. 
The fluctuation envelope, visualized by the shaded bands, highlights the strong QCM fluctuations at the separatrix and the positively skewed blob statistics in the SOL.
Validation extends from the means to the fluctuations.
Fig.~\ref{fig:profiles}(d) shows instantaneous $T_e$ profiles
whose variability agrees well with Thomson scattering measurements at different radii. 
QCM drives mesoscale radial oscillations of the pedestal foot across the separatrix in the simulation, consistent with the radial spread of the Thomson data at the separatrix.
This pattern is reminiscent of the experimental helium beam spectroscopy \cite{harrerQuasicontinuousExhaustScenario2022,grienerContinuousObservationFilaments2020}, which shows a coherent structure periodically crossing the separatrix.

A key experimental signature of the QCE regime is that the $T_e$ fall-off lengths $\lambda_{T_e}=|\partial_r \ln \bar{T_e}|^{-1}$ within and beyond the separatrix are decoupled. 
{This facilitates simultaneous high confinement and heat flux broadening} \cite{faitsch2021broadening} and is shown in Fig.~\ref{fig:profiles}(d), where $\lambda_{T_e}$ appears as linear slopes on a log-scale \cite{faitsch2021broadening,sun2015study}. 
Both experiment and simulation show a distinct SOL decay length (red dashed) that is clearly separated from the near-separatrix decay (blue dashed), the latter of which is linked to the presence of blobs.
We identify blobs by intermittency: localized, positively skewed density fluctuations exceeding $2.5$ standard deviations $\sigma$ \cite{manz2015origin}. 
Fig.~\ref{fig:profiles}(e) shows probability histograms of $\tilde{n}$ at the OMP. 
Near the separatrix, the QCM converts fluctuations from negatively skewed distributions for $\rpol<1$ to positively skewed distributions for $\rpol>1$.
Away from the separatrix, clear signatures of voids ($<-2.5\sigma$) occur for $\rpol<1$ and blobs ($>2.5\sigma$) for $\rpol>1$.
Notably, around $\rpol=1.01$ ($r-r_{\mathrm{sep}}=6.5\si{mm}$), where the QCM terminates, $\sigma$ drops sharply while the distribution retains a long positive tail. 
This radius coincides with the blob-launching surface in Fig.~\ref{fig:snaps} and with the beginning of $\lambda_{T_e}^{\mathrm{sol}}$ in Fig.~\ref{fig:profiles}(d).

\begin{figure} 
\centering
    \centering
    \includegraphics[trim=0.2cm 0.2cm 0.2cm 0.2cm, clip, width=0.496\linewidth]{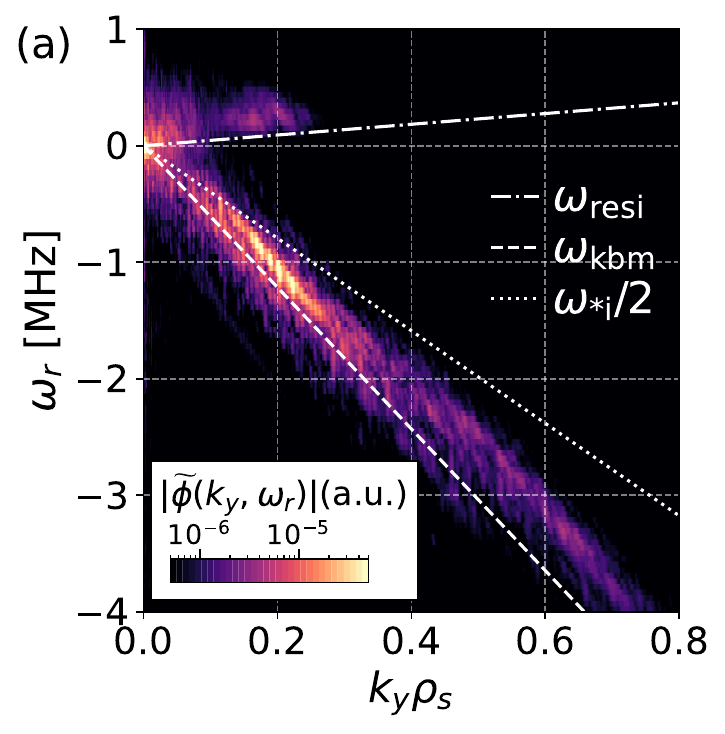}
    \includegraphics[trim=0.25cm 0.2cm 0.2cm 0.2cm, clip, width=0.492\linewidth]{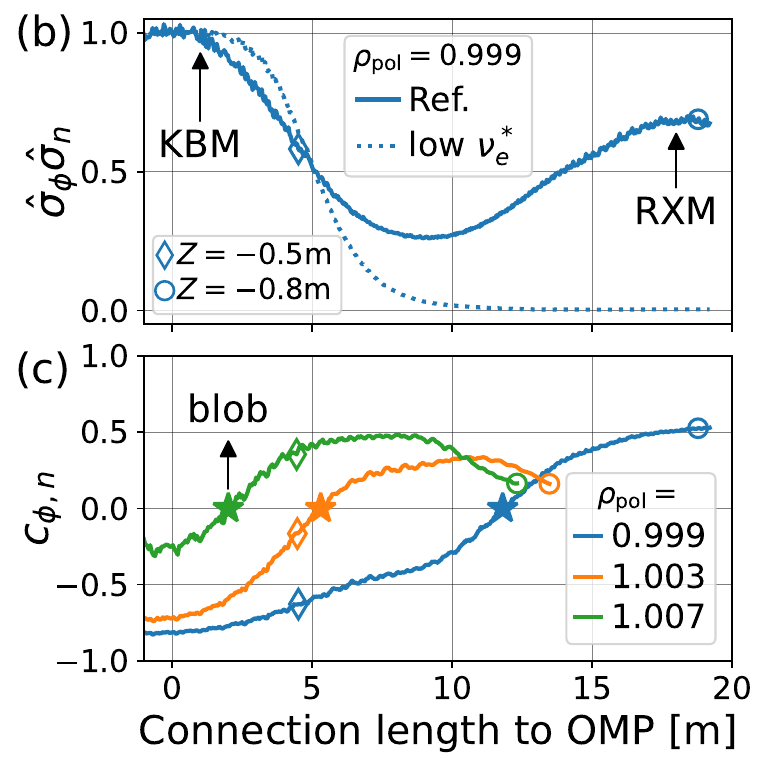}
    \vspace{-0.3cm}
    \caption{(a) Binormal Fourier spectra at $\rho_{\mathrm{pol}} = 0.999$ in the plasma frame. The dashed and dash–dotted lines indicate the KBM and RXM dispersion in \eqref{eqn:disp}, respectively. 
    (b) Production of fluctuations $\hat{\sigma}_\phi\hat{\sigma}_n$ ($\hat{\sigma}$ is the standard deviation normalized against its value at OMP) at $\rho_{\mathrm{pol}} = 0.999$ as a function of the parallel connection length to OMP, comparing Ref. and case C1 with lower collisionality.
    (c) Cross correlation between $\tilde{\phi}$ and $\tilde{n}$, plotted at different flux surfaces.
    $\lozenge$ and $\circ$ denote the heights $Z=-0.5\si{m}$ (where QCM reaches half-value) and $Z=-0.8\si{m}$ (near the X-point), respectively. \ding{73}  marks the zero-cross (where interchange occurs).}
    \label{fig:fourier}
\end{figure}

We now investigate the microinstabilities underlying the QCM and blob formation. Fig.~\ref{fig:fourier}(a) shows the real frequency $\omega_r$ versus binormal wavenumber $k_y$ at $\rpol=0.999$, evaluated in the plasma frame. The dominant branch propagates in the ion diamagnetic direction ($\omega_r<0$) and follows the theoretical kinetic ballooning mode (KBM) frequency $\omega_{\mathrm{kbm}}$. 
This branch also lies close to $\omega_{*i}/2$, consistent with the experimentally inferred QCM frequency scaling across multiple discharges \cite{kalis2023experimental}. 
A secondary branch propagates weakly in the electron direction ($\omega_r>0$) and follows the resistive-mode dispersion $\omega_{\mathrm{resi}}$. 
These theoretical dispersions follow from a linear analysis of the two-fluid model in \texttt{GRILLIX}:
\begin{equation}\label{eqn:disp}
\omega_{\mathrm{kbm}}=\frac{\frac{1}{6}+\eta_i}{1+\eta_i}\omega_{*i},\quad
\omega_{\mathrm{resi}} = \frac{\omega_{*e} + 0.71\eta_e\omega_* + \omega_{*i}}{3}\,.
\end{equation}
Here $\eta_i=L_n/L_{T_i}$, $\eta_e=L_n/L_{T_e}$, $L_n=|\partial_r \ln \bar{n} |^{-1}$, $L_{T_i}=|\partial_r \ln \bar{T_i}|^{-1}$, $L_{T_e}=|\partial_r \ln \bar{T_e}|^{-1}$. The diamagnetic frequency is $\omega_*=c_s k_y\rho_s/L_n$, with $\omega_{*i}=-(1+\eta_i)\omega_*$ and $\omega_{*e}=(1+\eta_e)\omega_*$. The KBM expression differs slightly from the $\omega_{*i}/2$ estimate reported in \cite{aleynikova2017quantitative} due to the inclusion of the time dependence of the ion diamagnetic velocity, $\partial_t\mathbf{u}_*$, in the polarization drift. And $\omega_{\mathrm{resi}}$ differs from the conventional resistive mode form \cite{ara1978magnetic,porcelli1986ion,kotschenreuther2019gyrokinetic,gupta2002violating,myra2005edge} due to the thermal-force contribution $0.71\eta_e\omega_*$ that is included in our model.
\eqref{eqn:disp} are obtained asymptotically by truncating higher-order terms in $L_n/R_0$ ($\approx 0.01$ in the simulation). 
In addition, $\omega_{\mathrm{kbm}}$ assumes the long-wavelength ordering $k_{\perp}^2\rho_s^2\ll 1$, whereas $\omega_{\mathrm{resi}}$ assumes strong magnetic diffusion, $\omega_{\eta}=\eta_\parallel k_{\perp}^2/\mu_0\gg\omega_*$ with $\eta_\parallel$ the parallel Spitzer resistivity. 
For these orderings constrain $k_\perp$ in opposite ways, they are most simultaneously satisfied in different regions: KBM is favored near the OMP, where $k_\perp\sim k_y$ is relatively small, while the resistive mode is favored closer to the X-point, where $k_\perp\gg k_y$ rises sharply due to the diverging connection length.
This argument is supported by the results in Fig.~\ref{fig:fourier}(b,c).
We first consider the surface $\rpol=0.999$.
The KBM, characterized by a kinetic-shear-Alfv\'en-wave-like \cite{doerk2016gyrokinetic} anti-correlation between $\tilde{\phi}$ and $\tilde{n}$ (the normalized covariance 
$c_{\phi,n}=\langle\tilde{\phi}\tilde{n}\rangle/(\langle\tilde{\phi}^2\rangle\langle\tilde{n}^2\rangle)^{1/2}<0$, where $\langle \circ \rangle$ denotes the time and toroidal average), dominates at OMP and is progressively stabilized by field-line bending when approaching the X-point. 
Resistivity is insufficient for a resistive ballooning mode to outcompete the KBM at the OMP. 
Yet resistive drive dominates near the X-point, where $\omega_\eta$ increases due to the rapid growth of $k_\perp$, exciting a resistive X-point mode (RXM) \cite{MyraDippolito2000,mazurenko2002experimental} with the drift-interchange nature $c_{\phi,n}>0$.

{Importantly, the QCM in Fig.~\ref{fig:snaps} is driven by the KBM rather than the RXM, while blobs are ejected when the RXM interacts with the KBM.}  
The highly shaped equilibrium, an access criterion for QCE, lengthens the parallel connection from OMP to the X-point region, making room for KBM and RXM to coexist at the same flux surface. 
Inevitably, KBM's and RXM's eigenfunctions with opposite signs of $c_{\phi,n}$ will meet and produce the interchange dynamics $c_{\phi,n}=0$ at some point (marked by \ding{73} in Fig.~\ref{fig:fourier}(c)).
At $\rpol=0.999$, the interchange couples close enough to the X-point to remain outside the poloidal extent of the QCM’s strong-fluctuation region (marked by $\lozenge$).
In the SOL, coupling strengthens as the connection length decreases and resistivity increases, shifting the interchange point toward the OMP as $\rpol$ rises.
By $\rpol=1.007$, the interchange lies within the QCM’s strong-fluctuation region.
Here, $E\times B$ advection efficiently expels the QCM’s positive density perturbations outward, launching the blobs seen in Fig.~\ref{fig:snaps} \cite{krasheninnikov2001scrape,d2011convective}. 
Consistently, a Fourier analysis of the phase shift,  $\alpha_{\phi,n}(k)=\mathrm{Arg}(\tilde{\phi}_k/\tilde{n}_k)$ with poloidal wavenumber $k$, peaks at the QCM poloidal scale near $\pi$ for $\rpol=0.999$ and gradually decreases to $\pi/2$ for $\rpol=1.007$. 

We substantiate the above interpretation with a comparative simulation.
Relative to the 
reference case, we repeat the run with identical parameters except for a reduced neutral density imposed at the divertor plates (case C1). 
This yields a lower $\nsep=3\times10^{19}/\si{m}^3$ in the saturated state. 
Case C1 exhibits a QCM similar to Fig.~\ref{fig:snaps}, but differs in four aspects. 
(1) The RXM branch is absent in the Fourier spectrum; only the KBM branch remains, with $\omega_r\approx 1.1\, \omega_{\mathrm{kbm}}$, higher harmonics, and stronger coherency. 
The dotted line in Fig.~\ref{fig:fourier}(b) confirms that KBM fluctuations decay toward the X-point without the emergence of another mode.
(2) The radial profile of the QCM phase shift drops very sharply from $\alpha_{\phi,n}\sim\pi$ to $\alpha_{\phi,n}\sim0$ at the outmost layer of QCM, without forming a robust interchange spectrum. 
(3) The SOL is dominated by continuous fluctuations with $\tilde{n}<2.5\sigma$, with no blob signatures, {despite a higher $E\times B$ shear rate in the SOL}. 
(4) $T_e$ exhibits a single exponential decay from the separatrix down to $20\,\si{eV}$ in the SOL, with no decoupling between $\lambda_{T_e}^{\mathrm{sep}}$ and $\lambda_{T_e}^{\mathrm{sol}}$.
Table~\ref{tab:sop} compares key separatrix parameters following \cite{faitsch2023analysis}: the MHD ballooning parameter $\alpha_m\sim \nsep T_{e,\mathrm{sep}}/\langle \lambda_p\rangle$  ($\langle \circ \rangle$ is the flux-surface average) and the edge collisionality $\nu^{*}_{e,\mathrm{O}}\sim \nsep/T_{e,\mathrm{sep}}^2$ measured at OMP, {and in addition, $\nu^{*}_{e,\mathrm{X}}$ measured at the orthogonal projection of the X-point onto flux surface $\rpol=0.999$.}
Both cases evolve to nearly identical $\alpha_m$, close to the ideal-ballooning threshold $\alpha_{c}$ that is calculated with \texttt{HELENA} for the magnetic equilibrium used here \cite{radovanovicDevelopingPhysicsUnderstanding2022}.
{Poloidally inhomogeneous collisionalities differentiate the two cases.
Particularly, a threefold lower $\nu^{*}_{e,\mathrm{X}}$ explains the absence of the RXM in case C1.}
Overall, case C1 resembles the Enhanced D-$\alpha$ regime in AUG, featuring a more coherent QCM but weaker blobs \cite{gil2020stationary}.

{The pedestal-foot transport is primarily set by the KBM nature of the QCM, rather than by a direct collisionality effect or by the RXM.}
Notably, KBM can also be unstable in the absence of a QCM during the inter-ELM phase \cite{zholobenko2024tokamak,snyder2011first,Dickinson2012}. 
To examine the difference between inter-ELM KBM and QCM, we include a third simulation C2 based on the inter-ELM phase of an ELMy H-mode equilibrium of AUG \#40411 \cite{zholobenko2024tokamak}.
In this case, neither the QCM nor blobs are observed, yet KBM remains at the pedestal foot. 
Despite similar $k_y\rho_s\approx 0.2$ and $\alpha_{\phi,n}\approx \pi$, the QCMs differ from the inter-ELM KBM by larger radial correlation length $L_c$ \cite{wessonTokamaks2011} in Table~\ref{tab:sop}. 
The $L_c$ of QCMs can even exceed the background pressure gradient length $\lambda_p= 8 \si{mm}$ at the separatrix, enabling mesoscopic heat-transport events from the pedestal to the SOL, in line with the pedestal oscillations in Fig.~\ref{fig:profiles}(d) and the streamer-like structures observed experimentally \cite{kalis2023experimental,grienerContinuousObservationFilaments2020,harrerQuasicontinuousExhaustScenario2022}.
The effective particle diffusivity $D^{\mathrm{eff}}$ and electron/ion heat conductivities $\chi_e^{\mathrm{eff}}$/$\chi_i^{\mathrm{eff}}$ are computed using a Fick-like relation \cite{callen2010analysis}. 
While Fick coefficients may be insufficient for describing non-local KBMs \cite{hahm2018mesoscopic,ashourvan2016mesoscopic}, they may serve as useful metrics radially inward where KBMs are first destabilized; these are reported in Table~\ref{tab:sop}. 
Transport in the QCM cases is substantially higher than inter-ELM KBM, corroborating the established view that sufficient transport by QCM limits pedestal widening and thereby avoids ELMs \cite{harrerParameterDependencesSmall2018,dunne2024quasi,snyder2011first}. 
{A detailed comparison yields the transport-strength ordering $\mathrm{C1}>\mathrm{Ref.}>\mathrm{C2}$, which clearly tracks $L_c$ rather than the collisionalities.}
In our model, $L_c$ is limited by $E\times B$ shear, shifting the focus to the radial electric field $E_r$.

\begin{table}[]
\centering
\begin{tabular}{c|cccccccccc}
& QCM & blob &\makecell{${\nsep}$} & $\alpha_{m}/\alpha_c$ & {$\nu^*_{e,\mathrm{O}}/\nu^*_{e,\mathrm{X}}$} 
& \makecell{$L_c$} 
& \makecell{$D^{\mathrm{eff}}$}  & 
\makecell{$\chi_e^{\mathrm{eff}}$}  
& \makecell{$\chi_i^{\mathrm{eff}}$}
\\ \hline 
{Ref.} 
& Yes & Yes & ${3.9}$ & $2.3/2.7$ & {${10}/28$} 
&$12$
& $0.27$ & $0.56$ & $1.2$  \\ 
{C1} 
& Yes & No & $3.0$ 
& $2.3/2.7$ 
& {$5.2/10$ }
& $17$
& $0.34$ & $0.70$ & $1.6$ \\ 
{C2} 
& No & No & $2.2$ & $1.2/2.9$ & {$5.3/7.1$} 
& $5.5$& $0.12$ & $0.23$ & $0.6$ \\ 
\end{tabular}
\caption{Comparison of key parameters for the reference case (Ref.), the case with same equilibrium but lower $\nsep$ (C1), and the case under ELMy equilibrium (C2). Here $\nsep$ is in $[10^{19}\, \si{m^{-3}}]$, and $L_c$ in $[\si{mm}]$. $D^{\mathrm{eff}}$, $\chi_e^{\mathrm{eff}}$, and $\chi_i^{\mathrm{eff}}$ in $[\si{m^2/s}]$ are measured at the start of the KBM-unstable $\rpol$.}
\label{tab:sop}
\end{table}
 
\begin{figure}[h]
\centering
    \centering
    \includegraphics[trim=0.2cm 0.2cm 0.2cm 0.2cm, clip, width=0.49\linewidth]{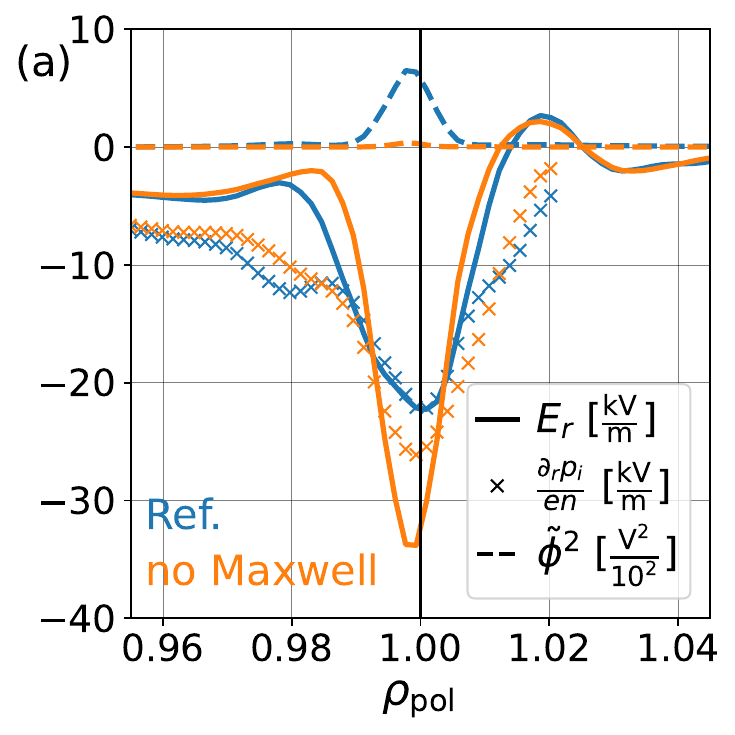}
    \includegraphics[trim=0.2cm 0.2cm 0.2cm 0.2cm, clip, width=0.48\linewidth]{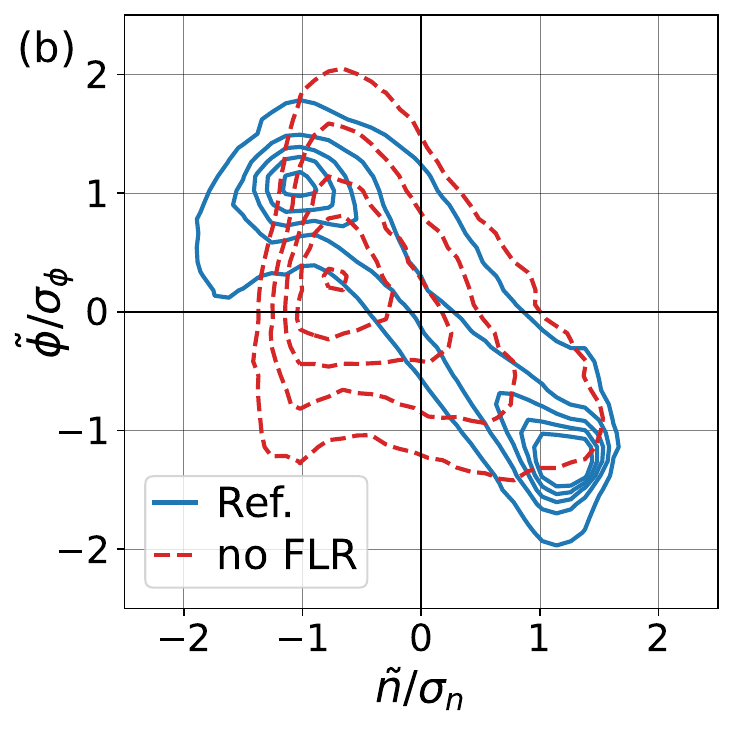}
    \caption{(a) Mean profiles of $E_r$, $(\partial_rp_i)/(en)$ and $\tilde{\phi}^2$, comparing the reference case and the test where the Maxwell stress is removed. (b) Cross coherence between $\tilde{\phi}$ and $\tilde{n}$, comparing the reference case and the test without FLR.}
    \label{fig:self}
\end{figure}

The role of $E_r$ in accessing QCE and avoiding Type-I ELMs has been noted in the experiments \cite{harrerQuasicontinuousExhaustScenario2022} and the MHD simulations \cite{cathey2022mhd}; here, we ground this role in first-principles dynamics underlying turbulence self-organization and transport regulation.
In our model, $E_r$ is not imposed but evolves through the vorticity dynamics. 
The radial force balance,
$E_r = ({\partial_r p_i})/({en}) + u_{\varphi} B_{\theta} -  u_{\theta} B_{\varphi} \,,$
is satisfied, where the ion pressure $p_i$ is shaped by transport, and the toroidal and poloidal flows  ($u_{\varphi}$, $u_{\theta}$) account for the neoclassical physics \cite{Rozhansky2009} and the turbulence effect \cite{diamond1991theory}. 
In turn, $E\times B$ shear feeds back on the QCM by regulating its $L_c$ and radial location.
In the reference simulation, the $E_r$ well at the OMP coincides radially with the peak of the fluctuation intensity of QCM (represented by the square of potential fluctuation $\tilde{\phi}^2$), as shown by the blue curves in Fig.~\ref{fig:self}. 
Near the separatrix, $E_r$ is dominated by $(\partial_r p_i)/(en)$, as the flow contributions $u_{\theta}B_{\varphi}$ and $u_{\varphi}B_{\theta}$ largely cancel. 
Given $E_r\approx(\partial_r p_i)/(en)$, a high $\nsep$ becomes a primary lever for flattening the $E_r$ well and establishing QCM with a large $L_c$.
This may largely explain why $\nsep$ appears as a main discriminator between QCE/EDA and ELMy operation in experiments \cite{faitsch2023analysis,miller2025determination}.
Additionally, at fixed $\nsep$, the estimation of $L_c$ is complicated by the $E_r\propto \partial_r p_i$ dependence.
The stiffness of $\partial_r p_i$ is determined by the transport whose saturation mechanism entails the zonal flow.
The zonally averaged poloidal flow $\langle u_{\theta}\rangle$ (zonal flow) is driven by the flow nonlinearity $\langle \tilde{u}_{r}\tilde{u}_{\theta}\rangle$ (Reynolds stress) \cite{smolyakov2000role,zholobenko2021electric,diamond2005zonal}.
At high $\alpha_m$, the zonal flow can be depleted by the magnetic nonlinearity $\langle \tilde{b}_{r} \tilde{b}_{\theta}\rangle$ (Maxwell stress) \cite{scott2005energetics,naulin2005shear}, which is found to be relevant to the present QCM cases.
The QCM efficiently produces Maxwell stress, characterised by the phase shift $\alpha_{{b}_{r},{b}_{\theta}}\approx 0$ at the QCM's wavenumber.
This Maxwell stress has a notable impact on the macroscopic profiles. This is confirmed by comparison with a control case derived from the reference case, in which the $\tilde{b}_{r}\tilde{b}_{\theta}$ nonlinearity is disabled, shown by orange curves in Fig.~\ref{fig:self}(a).
In this case, while $\nsep$ remains essentially unchanged, both $E_r$ and $\partial_rp_i$ evolve much deeper, and the QCM fluctuation vanishes. 
This outcome can be understood in two steps: (i) without Maxwell-stress depletion, the poloidal-flow contribution $-u_{\theta}B_{\varphi}$ exceeds the toroidal one, driving $E_r$ to deviate from $(\partial_r p_i)/(en)$; and (ii) the loss of QCM decreases the transport and steepens the $p_i$ pedestal, further deepening $(\partial_r p_i)/(en)$ and driving the system toward to an ELMy-H-mode-like state. 
We therefore conclude that Maxwell stress is a key component in the self-sustained QCM.

Despite the large fluctuations and long $L_c$, QCM-driven transport does not fully flatten the pedestal due to $\alpha_{\phi,n}\approx \pi$ (similarly for $\alpha_{\phi,T_e}$ and $\alpha_{\phi,T_i}$). 
This anti-correlated phase shift, also observed experimentally \cite{grenfell2024multi}, resembles kinetic-shear-Alfv\'en-wave oscillations \cite{scott2021turbulence} and reduces the efficiency of nonlinear transport. 
This behavior can be traced to the linear ballooning eigenproblem that governs the KBM,
which is derived from the two-fluid model used in \grillix{} and has the form 
\begin{equation} \label{eqn:kbm:sim_eqn}
    \frac{\mathrm{d}}{\mathrm{d} \theta } f_{\theta}\frac{\mathrm{d}\Phi}{\mathrm{d} \theta } = \bigg( 
    C_{0,\theta} + C_{2,\theta} {\omega}^2+ C_{1,\theta} {\omega}+\frac{C_{-1,\theta}}{{\omega}-{\omega}_1}\bigg)\Phi\,,
\end{equation}
where $\Phi={\tilde{A}_{\parallel,\mathbf{k}}}/({k_\parallel R_0 })$ is the Fourier eigenfunction to be solved, $\omega=\omega_r+\gamma i$ is the complex eigenfrequency and $\theta$ is the ballooning angle~\cite{connor1978shear}. $f_{\theta}=1+(s \theta-\alpha_m \sin \theta)^2$ is a function of $\theta$ depending on $\alpha_m$ and the magnetic shear $s$.
Inside the bracket is a polynomial of $\omega^j$ with the prefactor $C_{j,\theta}$, which assembles the key physics determining the instability of the system.
The $C_{0,\theta}$ and $C_{2,\theta}$ terms recover the ideal ballooning mode \cite{connor1978shear},
which has a quadratic form of $\omega$ and hence $\omega_r=0$ and the approximated interchange-like phase shift $\alpha_{\phi,n} \approx \arg(\omega) = \pi/2$.
The $C_{-1,\theta}$ term represents the ion magnetic drift resonance, originating from the ion's perpendicular heat conduction as a two-fluid effect in toroidal geometry, and destabilizes the system by reducing $\alpha_m$ threshold given a high $\eta_i$.
Importantly, the $C_{1,\theta}$ term occurs when the diamagnetic polarization, i.e.~finite Larmor radius (FLR) term $\nabla\cdot (\partial_t+\mathbf{u}_{E} \cdot \nabla ) \mathbf{u}_{*}$ is retained in the vorticity equation.
Its inclusion introduces a significant real frequency component $\omega_r=\omega_{\mathrm{kbm}}<0$ consistent with \cite{aleynikova2017quantitative}.
Consequently, when KBM is marginally unstable, we expect $\alpha_{\phi,n}\approx \pi$ for $\gamma \ll \omega_{\mathrm{kbm}}$. 
The significance of FLR physics in the KBM turbulence is further substantiated by performing a test branched from the reference case, wherein we disable the term $\nabla\cdot(\mathbf{u}_{E} \cdot \nabla \mathbf{u}_{*})$.
Fig.~\ref{fig:self}(b) compares the cross-coherence between $\tilde{\phi}$ and $\tilde{n}$ at $\rpol=0.999$ near the OMP for the reference case and the no-FLR test. 
The probability density increases linearly from the outer to the inner contours. 
When FLR is removed, the picture shifts from a coherent pattern aligned diagonally  ($\alpha_{\phi,n}\approx \pi$) to broadband turbulence aligned horizontally ($\alpha_{\phi,n}\approx \pi/2$). 
The no-FLR case, subject to interchange-like dynamics, exhibits a tenfold increase in $E\times B$ heat flux, leading to pedestal collapse and thus discrepancy with the experiment.
These results underscore that FLR stabilization is essential for sustaining the coherent QCM structure with $\alpha_{\phi,n}\approx \pi$, and for preventing its transport from degrading the confinement.

{To summarize, validated simulations reveal the turbulent nature of the Quasicontinuous Exhaust regime, a promising scenario for fusion reactors.
The experimentally observed QCM and blob dynamics, reproduced in the simulations, are governed by two separatrix-spanning modes.
The QCM is a kinetic ballooning mode (KBM) that develops an extended, mesoscopic radial correlation length via electromagnetic self-organization of turbulence, thereby driving enhanced transport.
Blob launching occurs via interchange dynamics when this upstream KBM activity interacts with a resistive mode originating from the X-point region (RXM).}

{This work provides a sound physics basis for extrapolations to future reactors. 
Our results suggest, first, that access to the KBM-dominated QCM should depend primarily on the beta-related ballooning parameter \cite{dunne2024quasi,faitsch2025quasi}, rather than on collisionality-related parameters \cite{Zeiler1998,scott2005drift,eich2020turbulence} as assumed in extrapolations based on resistivity-dominated modes \cite{faitsch2023analysis}.
However, the experimentally observed separatrix density dependence \cite{faitsch2023analysis,miller2025determination} may instead reflect the requirement for reduced $E\times B$ shear.
Quantitatively predicting whether QCM transport is sufficient to suppress ELMs, yet not so excessive as to degrade confinement, will require parameterizing the $E\times B$ shear weakening by Maxwell stress and interchange stabilization by finite-Larmor-radius physics, respectively. 
Second, blob-enhanced exhaust requires the coexistence of those KBMs with an RXM destabilized by elevated local collisionality near the X-point, rather than the excitation of a resistive ballooning mode by high upstream collisionality.
Thus, despite the more constrained access to high outboard-midplane separatrix collisionality in reactor-scale devices such as ITER \cite{labit2019dependence,faitsch2025quasi}, the mechanisms elucidated herein may nevertheless remain applicable for reconciling high confinement with effective heat exhaust in future fusion plasmas.
} 

\section*{Acknowledgements}
The authors would like to thank P. Manz, T. Pütterich, G. Birkenmeier, {M. Dunne, A. Cathey}, P. Ulbl, J. Kalis, J. Pfennig, L. Radovanovic, and B. Frei for fruitful discussions on many intricate topics.
This work was carried out within the framework of the EUROfusion Consortium, funded by the European Union via the Euratom Research and Training Programme (Grant Agreement No 101052200 — EUROfusion). 
Views and opinions expressed are however those of the author(s) only and do not necessarily reflect those of the European Union or the European Commission. Neither the European Union nor the European Commission can be held responsible for them. 
The simulations were performed on the national supercomputer HPE Apollo Hawk at the High Performance Computing Center Stuttgart (HLRS) under the grant number GRILLIX/44281.
\input{myrefs.bbl}
\end{document}

%% file: myrefs.bbl
%

%% file: main.bbl
\begin{thebibliography}{66}%
\makeatletter
\providecommand \@ifxundefined [1]{%
 \@ifx{#1\undefined}
}%
\providecommand \@ifnum [1]{%
 \ifnum #1\expandafter \@firstoftwo
 \else \expandafter \@secondoftwo
 \fi
}%
\providecommand \@ifx [1]{%
 \ifx #1\expandafter \@firstoftwo
 \else \expandafter \@secondoftwo
 \fi
}%
\providecommand \natexlab [1]{#1}%
\providecommand \enquote  [1]{``#1''}%
\providecommand \bibnamefont  [1]{#1}%
\providecommand \bibfnamefont [1]{#1}%
\providecommand \citenamefont [1]{#1}%
\providecommand \href@noop [0]{\@secondoftwo}%
\providecommand \href [0]{\begingroup \@sanitize@url \@href}%
\providecommand \@href[1]{\@@startlink{#1}\@@href}%
\providecommand \@@href[1]{\endgroup#1\@@endlink}%
\providecommand \@sanitize@url [0]{\catcode `\\12\catcode `\$12\catcode
  `\&12\catcode `\#12\catcode `\^12\catcode `\_12\catcode `\%12\relax}%
\providecommand \@@startlink[1]{}%
\providecommand \@@endlink[0]{}%
\providecommand \url  [0]{\begingroup\@sanitize@url \@url }%
\providecommand \@url [1]{\endgroup\@href {#1}{\urlprefix }}%
\providecommand \urlprefix  [0]{URL }%
\providecommand \Eprint [0]{\href }%
\providecommand \doibase [0]{https://doi.org/}%
\providecommand \selectlanguage [0]{\@gobble}%
\providecommand \bibinfo  [0]{\@secondoftwo}%
\providecommand \bibfield  [0]{\@secondoftwo}%
\providecommand \translation [1]{[#1]}%
\providecommand \BibitemOpen [0]{}%
\providecommand \bibitemStop [0]{}%
\providecommand \bibitemNoStop [0]{.\EOS\space}%
\providecommand \EOS [0]{\spacefactor3000\relax}%
\providecommand \BibitemShut  [1]{\csname bibitem#1\endcsname}%
\let\auto@bib@innerbib\@empty
\bibitem [{\citenamefont {Siccinio}\ \emph {et~al.}(2020)\citenamefont
  {Siccinio}, \citenamefont {Biel}, \citenamefont {Cavedon} \emph
  {et~al.}}]{siccinio2020demo}%
  \BibitemOpen
  \bibfield  {author} {\bibinfo {author} {\bibfnamefont {M.}~\bibnamefont
  {Siccinio}}, \bibinfo {author} {\bibfnamefont {W.}~\bibnamefont {Biel}},
  \bibinfo {author} {\bibfnamefont {M.}~\bibnamefont {Cavedon}}, \emph
  {et~al.},\ }\href {https://doi.org/10.1016/j.fusengdes.2020.111603}
  {\bibfield  {journal} {\bibinfo  {journal} {Fusion Engineering and Design}\
  }\textbf {\bibinfo {volume} {156}},\ \bibinfo {pages} {111603} (\bibinfo
  {year} {2020})}\BibitemShut {NoStop}%
\bibitem [{\citenamefont {Viezzer}\ and\ \citenamefont
  {et~al.}(2018)}]{Viezzer2018}%
  \BibitemOpen
  \bibfield  {author} {\bibinfo {author} {\bibfnamefont {E.}~\bibnamefont
  {Viezzer}}\ and\ \bibinfo {author} {\bibnamefont {et~al.}},\ }\href
  {https://doi.org/10.1088/1741-4326/aacc6c} {\bibfield  {journal} {\bibinfo
  {journal} {Nuclear Fusion}\ }\textbf {\bibinfo {volume} {58}},\ \bibinfo
  {pages} {082017} (\bibinfo {year} {2018})}\BibitemShut {NoStop}%
\bibitem [{\citenamefont {Harrer}\ \emph {et~al.}(2018)\citenamefont {Harrer},
  \citenamefont {Wolfrum}, \citenamefont {Dunne}, \citenamefont {Manz},
  \citenamefont {Cavedon}, \citenamefont {Lang}, \citenamefont {Kurzan},
  \citenamefont {Eich}, \citenamefont {Labit}, \citenamefont {Stober} \emph
  {et~al.}}]{harrerParameterDependencesSmall2018}%
  \BibitemOpen
  \bibfield  {author} {\bibinfo {author} {\bibfnamefont {G.}~\bibnamefont
  {Harrer}}, \bibinfo {author} {\bibfnamefont {E.}~\bibnamefont {Wolfrum}},
  \bibinfo {author} {\bibfnamefont {M.}~\bibnamefont {Dunne}}, \bibinfo
  {author} {\bibfnamefont {P.}~\bibnamefont {Manz}}, \bibinfo {author}
  {\bibfnamefont {M.}~\bibnamefont {Cavedon}}, \bibinfo {author} {\bibfnamefont
  {P.}~\bibnamefont {Lang}}, \bibinfo {author} {\bibfnamefont {B.}~\bibnamefont
  {Kurzan}}, \bibinfo {author} {\bibfnamefont {T.}~\bibnamefont {Eich}},
  \bibinfo {author} {\bibfnamefont {B.}~\bibnamefont {Labit}}, \bibinfo
  {author} {\bibfnamefont {J.}~\bibnamefont {Stober}}, \emph {et~al.},\ }\href
  {https://doi.org/10.1088/1741-4326/aad757} {\bibfield  {journal} {\bibinfo
  {journal} {Nuclear Fusion}\ }\textbf {\bibinfo {volume} {58}},\ \bibinfo
  {pages} {112001} (\bibinfo {year} {2018})}\BibitemShut {NoStop}%
\bibitem [{\citenamefont {Labit}\ \emph {et~al.}(2019)\citenamefont {Labit},
  \citenamefont {Eich}, \citenamefont {Harrer}, \citenamefont {Wolfrum},
  \citenamefont {Bernert}, \citenamefont {Dunne}, \citenamefont {Frassinetti},
  \citenamefont {Hennequin}, \citenamefont {Maurizio}, \citenamefont {Merle}
  \emph {et~al.}}]{labit2019dependence}%
  \BibitemOpen
  \bibfield  {author} {\bibinfo {author} {\bibfnamefont {B.}~\bibnamefont
  {Labit}}, \bibinfo {author} {\bibfnamefont {T.}~\bibnamefont {Eich}},
  \bibinfo {author} {\bibfnamefont {G.~F.}\ \bibnamefont {Harrer}}, \bibinfo
  {author} {\bibfnamefont {E.}~\bibnamefont {Wolfrum}}, \bibinfo {author}
  {\bibfnamefont {M.}~\bibnamefont {Bernert}}, \bibinfo {author} {\bibfnamefont
  {M.}~\bibnamefont {Dunne}}, \bibinfo {author} {\bibfnamefont
  {L.}~\bibnamefont {Frassinetti}}, \bibinfo {author} {\bibfnamefont
  {P.}~\bibnamefont {Hennequin}}, \bibinfo {author} {\bibfnamefont
  {R.}~\bibnamefont {Maurizio}}, \bibinfo {author} {\bibfnamefont
  {A.}~\bibnamefont {Merle}}, \emph {et~al.},\ }\href
  {https://doi.org/10.1088/1741-4326/ab2211} {\bibfield  {journal} {\bibinfo
  {journal} {Nuclear fusion}\ }\textbf {\bibinfo {volume} {59}},\ \bibinfo
  {pages} {086020} (\bibinfo {year} {2019})}\BibitemShut {NoStop}%
\bibitem [{\citenamefont {Faitsch}\ \emph {et~al.}(2025)\citenamefont
  {Faitsch}, \citenamefont {Dunne}, \citenamefont {Lerche}, \citenamefont
  {Lomas}, \citenamefont {Balboa}, \citenamefont {Bilkova}, \citenamefont
  {Bohm}, \citenamefont {Kappatou}, \citenamefont {Kos}, \citenamefont {Labit}
  \emph {et~al.}}]{faitsch2025quasi}%
  \BibitemOpen
  \bibfield  {author} {\bibinfo {author} {\bibfnamefont {M.}~\bibnamefont
  {Faitsch}}, \bibinfo {author} {\bibfnamefont {M.}~\bibnamefont {Dunne}},
  \bibinfo {author} {\bibfnamefont {E.}~\bibnamefont {Lerche}}, \bibinfo
  {author} {\bibfnamefont {P.}~\bibnamefont {Lomas}}, \bibinfo {author}
  {\bibfnamefont {I.}~\bibnamefont {Balboa}}, \bibinfo {author} {\bibfnamefont
  {P.}~\bibnamefont {Bilkova}}, \bibinfo {author} {\bibfnamefont
  {P.}~\bibnamefont {Bohm}}, \bibinfo {author} {\bibfnamefont {A.}~\bibnamefont
  {Kappatou}}, \bibinfo {author} {\bibfnamefont {D.}~\bibnamefont {Kos}},
  \bibinfo {author} {\bibfnamefont {B.}~\bibnamefont {Labit}}, \emph {et~al.},\
  }\href {https://doi.org/10.1088/1741-4326/adaa86} {\bibfield  {journal}
  {\bibinfo  {journal} {Nuclear Fusion}\ }\textbf {\bibinfo {volume} {65}},\
  \bibinfo {pages} {024003} (\bibinfo {year} {2025})}\BibitemShut {NoStop}%
\bibitem [{\citenamefont {Dunne}\ \emph {et~al.}(2024)\citenamefont {Dunne},
  \citenamefont {Faitsch}, \citenamefont {Radovanovic}, \citenamefont
  {Wolfrum}, \citenamefont {Team} \emph {et~al.}}]{dunne2024quasi}%
  \BibitemOpen
  \bibfield  {author} {\bibinfo {author} {\bibfnamefont {M.}~\bibnamefont
  {Dunne}}, \bibinfo {author} {\bibfnamefont {M.}~\bibnamefont {Faitsch}},
  \bibinfo {author} {\bibfnamefont {L.}~\bibnamefont {Radovanovic}}, \bibinfo
  {author} {\bibfnamefont {E.}~\bibnamefont {Wolfrum}}, \bibinfo {author}
  {\bibfnamefont {A.~U.}\ \bibnamefont {Team}}, \emph {et~al.},\ }\href
  {https://doi.org/10.1088/1741-4326/ad89da} {\bibfield  {journal} {\bibinfo
  {journal} {Nuclear Fusion}\ }\textbf {\bibinfo {volume} {64}},\ \bibinfo
  {pages} {124003} (\bibinfo {year} {2024})}\BibitemShut {NoStop}%
\bibitem [{\citenamefont {Faitsch}\ \emph {et~al.}(2023)\citenamefont
  {Faitsch}, \citenamefont {Eich}, \citenamefont {Harrer}, \citenamefont
  {Wolfrum}, \citenamefont {Brida}, \citenamefont {David}, \citenamefont
  {Dunne}, \citenamefont {Gil}, \citenamefont {Labit}, \citenamefont {Stroth}
  \emph {et~al.}}]{faitsch2023analysis}%
  \BibitemOpen
  \bibfield  {author} {\bibinfo {author} {\bibfnamefont {M.}~\bibnamefont
  {Faitsch}}, \bibinfo {author} {\bibfnamefont {T.}~\bibnamefont {Eich}},
  \bibinfo {author} {\bibfnamefont {G.}~\bibnamefont {Harrer}}, \bibinfo
  {author} {\bibfnamefont {E.}~\bibnamefont {Wolfrum}}, \bibinfo {author}
  {\bibfnamefont {D.}~\bibnamefont {Brida}}, \bibinfo {author} {\bibfnamefont
  {P.}~\bibnamefont {David}}, \bibinfo {author} {\bibfnamefont
  {M.}~\bibnamefont {Dunne}}, \bibinfo {author} {\bibfnamefont
  {L.}~\bibnamefont {Gil}}, \bibinfo {author} {\bibfnamefont {B.}~\bibnamefont
  {Labit}}, \bibinfo {author} {\bibfnamefont {U.}~\bibnamefont {Stroth}}, \emph
  {et~al.},\ }\href {https://doi.org/10.1088/1741-4326/acd464} {\bibfield
  {journal} {\bibinfo  {journal} {Nuclear Fusion}\ }\textbf {\bibinfo {volume}
  {63}},\ \bibinfo {pages} {076013} (\bibinfo {year} {2023})}\BibitemShut
  {NoStop}%
\bibitem [{\citenamefont {Wagner}\ \emph {et~al.}(1982)\citenamefont {Wagner},
  \citenamefont {Becker}, \citenamefont {Behringer} \emph
  {et~al.}}]{wagner1982regime}%
  \BibitemOpen
  \bibfield  {author} {\bibinfo {author} {\bibfnamefont {F.}~\bibnamefont
  {Wagner}}, \bibinfo {author} {\bibfnamefont {G.}~\bibnamefont {Becker}},
  \bibinfo {author} {\bibfnamefont {K.}~\bibnamefont {Behringer}}, \emph
  {et~al.},\ }\href {https://doi.org/10.1103/PhysRevLett.49.1408} {\bibfield
  {journal} {\bibinfo  {journal} {Physical Review Letters}\ }\textbf {\bibinfo
  {volume} {49}},\ \bibinfo {pages} {1408} (\bibinfo {year}
  {1982})}\BibitemShut {NoStop}%
\bibitem [{\citenamefont {Zohm}(1996)}]{zohm1996edge}%
  \BibitemOpen
  \bibfield  {author} {\bibinfo {author} {\bibfnamefont {H.}~\bibnamefont
  {Zohm}},\ }\href {https://doi.org/10.1088/0741-3335/38/2/001} {\bibfield
  {journal} {\bibinfo  {journal} {Plasma Physics and Controlled Fusion}\
  }\textbf {\bibinfo {volume} {38}},\ \bibinfo {pages} {105} (\bibinfo {year}
  {1996})}\BibitemShut {NoStop}%
\bibitem [{\citenamefont {Eich}\ and\ \citenamefont {et~al.}(2013)}]{Eich2013}%
  \BibitemOpen
  \bibfield  {author} {\bibinfo {author} {\bibfnamefont {T.}~\bibnamefont
  {Eich}}\ and\ \bibinfo {author} {\bibnamefont {et~al.}},\ }\href
  {https://doi.org/10.1088/0029-5515/53/9/093031} {\bibfield  {journal}
  {\bibinfo  {journal} {Nuclear Fusion}\ }\textbf {\bibinfo {volume} {53}},\
  \bibinfo {pages} {093031} (\bibinfo {year} {2013})}\BibitemShut {NoStop}%
\bibitem [{\citenamefont {Faitsch}\ \emph {et~al.}(2021)\citenamefont
  {Faitsch}, \citenamefont {Eich}, \citenamefont {Harrer}, \citenamefont
  {Wolfrum}, \citenamefont {Brida}, \citenamefont {David}, \citenamefont
  {Griener}, \citenamefont {Stroth}, \citenamefont {Team}, \citenamefont {Team}
  \emph {et~al.}}]{faitsch2021broadening}%
  \BibitemOpen
  \bibfield  {author} {\bibinfo {author} {\bibfnamefont {M.}~\bibnamefont
  {Faitsch}}, \bibinfo {author} {\bibfnamefont {T.}~\bibnamefont {Eich}},
  \bibinfo {author} {\bibfnamefont {G.}~\bibnamefont {Harrer}}, \bibinfo
  {author} {\bibfnamefont {E.}~\bibnamefont {Wolfrum}}, \bibinfo {author}
  {\bibfnamefont {D.}~\bibnamefont {Brida}}, \bibinfo {author} {\bibfnamefont
  {P.}~\bibnamefont {David}}, \bibinfo {author} {\bibfnamefont
  {M.}~\bibnamefont {Griener}}, \bibinfo {author} {\bibfnamefont
  {U.}~\bibnamefont {Stroth}}, \bibinfo {author} {\bibfnamefont {A.~U.}\
  \bibnamefont {Team}}, \bibinfo {author} {\bibfnamefont {E.~M.}\ \bibnamefont
  {Team}}, \emph {et~al.},\ }\href {https://doi.org/10.1016/j.nme.2020.100890}
  {\bibfield  {journal} {\bibinfo  {journal} {Nuclear Materials and Energy}\
  }\textbf {\bibinfo {volume} {26}},\ \bibinfo {pages} {100890} (\bibinfo
  {year} {2021})}\BibitemShut {NoStop}%
\bibitem [{\citenamefont {Wischmeier}\ \emph {et~al.}(2015)\citenamefont
  {Wischmeier}, \citenamefont {Team} \emph {et~al.}}]{wischmeier2015high}%
  \BibitemOpen
  \bibfield  {author} {\bibinfo {author} {\bibfnamefont {M.}~\bibnamefont
  {Wischmeier}}, \bibinfo {author} {\bibfnamefont {A.~U.}\ \bibnamefont
  {Team}}, \emph {et~al.},\ }\href
  {https://doi.org/10.1016/j.jnucmat.2014.12.078} {\bibfield  {journal}
  {\bibinfo  {journal} {Journal of Nuclear Materials}\ }\textbf {\bibinfo
  {volume} {463}},\ \bibinfo {pages} {22} (\bibinfo {year} {2015})}\BibitemShut
  {NoStop}%
\bibitem [{\citenamefont {Mazurenko}\ \emph {et~al.}(2002)\citenamefont
  {Mazurenko}, \citenamefont {Porkolab}, \citenamefont {Mossessian},
  \citenamefont {Snipes}, \citenamefont {Xu},\ and\ \citenamefont
  {Nevins}}]{mazurenko2002experimental}%
  \BibitemOpen
  \bibfield  {author} {\bibinfo {author} {\bibfnamefont {A.}~\bibnamefont
  {Mazurenko}}, \bibinfo {author} {\bibfnamefont {M.}~\bibnamefont {Porkolab}},
  \bibinfo {author} {\bibfnamefont {D.}~\bibnamefont {Mossessian}}, \bibinfo
  {author} {\bibfnamefont {J.}~\bibnamefont {Snipes}}, \bibinfo {author}
  {\bibfnamefont {X.}~\bibnamefont {Xu}},\ and\ \bibinfo {author}
  {\bibfnamefont {W.}~\bibnamefont {Nevins}},\ }\href
  {https://doi.org/10.1103/PhysRevLett.89.225004} {\bibfield  {journal}
  {\bibinfo  {journal} {Physical Review Letters}\ }\textbf {\bibinfo {volume}
  {89}},\ \bibinfo {pages} {225004} (\bibinfo {year} {2002})}\BibitemShut
  {NoStop}%
\bibitem [{\citenamefont {Kalis}\ \emph {et~al.}(2023)\citenamefont {Kalis},
  \citenamefont {Birkenmeier}, \citenamefont {Manz}, \citenamefont {Eich},
  \citenamefont {Griener}, \citenamefont {Goti}, \citenamefont {Cavedon},
  \citenamefont {Gil}, \citenamefont {Faitsch}, \citenamefont {Radovanovic}
  \emph {et~al.}}]{kalis2023experimental}%
  \BibitemOpen
  \bibfield  {author} {\bibinfo {author} {\bibfnamefont {J.}~\bibnamefont
  {Kalis}}, \bibinfo {author} {\bibfnamefont {G.}~\bibnamefont {Birkenmeier}},
  \bibinfo {author} {\bibfnamefont {P.}~\bibnamefont {Manz}}, \bibinfo {author}
  {\bibfnamefont {T.}~\bibnamefont {Eich}}, \bibinfo {author} {\bibfnamefont
  {M.}~\bibnamefont {Griener}}, \bibinfo {author} {\bibfnamefont
  {R.}~\bibnamefont {Goti}}, \bibinfo {author} {\bibfnamefont {M.}~\bibnamefont
  {Cavedon}}, \bibinfo {author} {\bibfnamefont {L.}~\bibnamefont {Gil}},
  \bibinfo {author} {\bibfnamefont {M.}~\bibnamefont {Faitsch}}, \bibinfo
  {author} {\bibfnamefont {L.}~\bibnamefont {Radovanovic}}, \emph {et~al.},\
  }\href {https://doi.org/10.1088/1741-4326/ad0d32} {\bibfield  {journal}
  {\bibinfo  {journal} {Nuclear Fusion}\ }\textbf {\bibinfo {volume} {64}},\
  \bibinfo {pages} {016038} (\bibinfo {year} {2023})}\BibitemShut {NoStop}%
\bibitem [{\citenamefont {Terry}\ \emph {et~al.}(2005)\citenamefont {Terry},
  \citenamefont {Basse}, \citenamefont {Cziegler}, \citenamefont {Greenwald},
  \citenamefont {Grulke}, \citenamefont {LaBombard}, \citenamefont {Zweben},
  \citenamefont {Edlund}, \citenamefont {Hughes}, \citenamefont {Lin} \emph
  {et~al.}}]{terry2005transport}%
  \BibitemOpen
  \bibfield  {author} {\bibinfo {author} {\bibfnamefont {J.}~\bibnamefont
  {Terry}}, \bibinfo {author} {\bibfnamefont {N.}~\bibnamefont {Basse}},
  \bibinfo {author} {\bibfnamefont {I.}~\bibnamefont {Cziegler}}, \bibinfo
  {author} {\bibfnamefont {M.}~\bibnamefont {Greenwald}}, \bibinfo {author}
  {\bibfnamefont {O.}~\bibnamefont {Grulke}}, \bibinfo {author} {\bibfnamefont
  {B.}~\bibnamefont {LaBombard}}, \bibinfo {author} {\bibfnamefont
  {S.}~\bibnamefont {Zweben}}, \bibinfo {author} {\bibfnamefont
  {E.}~\bibnamefont {Edlund}}, \bibinfo {author} {\bibfnamefont
  {J.}~\bibnamefont {Hughes}}, \bibinfo {author} {\bibfnamefont
  {L.}~\bibnamefont {Lin}}, \emph {et~al.},\ }\href
  {https://doi.org/10.1088/0029-5515/45/11/013} {\bibfield  {journal} {\bibinfo
   {journal} {Nuclear Fusion}\ }\textbf {\bibinfo {volume} {45}},\ \bibinfo
  {pages} {1321} (\bibinfo {year} {2005})}\BibitemShut {NoStop}%
\bibitem [{\citenamefont {D’ippolito}\ \emph {et~al.}(2011)\citenamefont
  {D’ippolito}, \citenamefont {Myra},\ and\ \citenamefont
  {Zweben}}]{d2011convective}%
  \BibitemOpen
  \bibfield  {author} {\bibinfo {author} {\bibfnamefont {D.}~\bibnamefont
  {D’ippolito}}, \bibinfo {author} {\bibfnamefont {J.}~\bibnamefont {Myra}},\
  and\ \bibinfo {author} {\bibfnamefont {S.}~\bibnamefont {Zweben}},\
  }\bibfield  {journal} {\bibinfo  {journal} {Physics of Plasmas}\ }\textbf
  {\bibinfo {volume} {18}},\ \href {https://doi.org/10.1063/1.3594609}
  {10.1063/1.3594609} (\bibinfo {year} {2011})\BibitemShut {NoStop}%
\bibitem [{\citenamefont {Stegmeir}\ \emph {et~al.}(2019)\citenamefont
  {Stegmeir}, \citenamefont {Ross}, \citenamefont {Body}, \citenamefont
  {Francisquez}, \citenamefont {Zholobenko}, \citenamefont {Coster},
  \citenamefont {Maj}, \citenamefont {Manz}, \citenamefont {Jenko},
  \citenamefont {Rogers} \emph {et~al.}}]{stegmeir2019global}%
  \BibitemOpen
  \bibfield  {author} {\bibinfo {author} {\bibfnamefont {A.}~\bibnamefont
  {Stegmeir}}, \bibinfo {author} {\bibfnamefont {A.}~\bibnamefont {Ross}},
  \bibinfo {author} {\bibfnamefont {T.}~\bibnamefont {Body}}, \bibinfo {author}
  {\bibfnamefont {M.}~\bibnamefont {Francisquez}}, \bibinfo {author}
  {\bibfnamefont {W.}~\bibnamefont {Zholobenko}}, \bibinfo {author}
  {\bibfnamefont {D.}~\bibnamefont {Coster}}, \bibinfo {author} {\bibfnamefont
  {O.}~\bibnamefont {Maj}}, \bibinfo {author} {\bibfnamefont {P.}~\bibnamefont
  {Manz}}, \bibinfo {author} {\bibfnamefont {F.}~\bibnamefont {Jenko}},
  \bibinfo {author} {\bibfnamefont {B.}~\bibnamefont {Rogers}}, \emph
  {et~al.},\ }\href {https://doi.org/10.1063/1.5089864} {\bibfield  {journal}
  {\bibinfo  {journal} {Physics of Plasmas}\ }\textbf {\bibinfo {volume}
  {26}},\ \bibinfo {pages} {052517} (\bibinfo {year} {2019})}\BibitemShut
  {NoStop}%
\bibitem [{\citenamefont {Zhang}\ \emph {et~al.}(2024)\citenamefont {Zhang},
  \citenamefont {Zholobenko}, \citenamefont {Stegmeir}, \citenamefont {Eder},\
  and\ \citenamefont {Jenko}}]{zhang2024magnetic}%
  \BibitemOpen
  \bibfield  {author} {\bibinfo {author} {\bibfnamefont {K.}~\bibnamefont
  {Zhang}}, \bibinfo {author} {\bibfnamefont {W.}~\bibnamefont {Zholobenko}},
  \bibinfo {author} {\bibfnamefont {A.}~\bibnamefont {Stegmeir}}, \bibinfo
  {author} {\bibfnamefont {K.}~\bibnamefont {Eder}},\ and\ \bibinfo {author}
  {\bibfnamefont {F.}~\bibnamefont {Jenko}},\ }\href
  {https://doi.org/10.1088/1741-4326/ad1b93} {\bibfield  {journal} {\bibinfo
  {journal} {Nuclear Fusion}\ }\textbf {\bibinfo {volume} {64}},\ \bibinfo
  {pages} {036016} (\bibinfo {year} {2024})}\BibitemShut {NoStop}%
\bibitem [{\citenamefont {Zhang}\ \emph {et~al.}(2025)\citenamefont {Zhang},
  \citenamefont {Zholobenko}, \citenamefont {Stegmeir}, \citenamefont {Eder},\
  and\ \citenamefont {Jenko}}]{zhang2025dynamical}%
  \BibitemOpen
  \bibfield  {author} {\bibinfo {author} {\bibfnamefont {K.}~\bibnamefont
  {Zhang}}, \bibinfo {author} {\bibfnamefont {W.}~\bibnamefont {Zholobenko}},
  \bibinfo {author} {\bibfnamefont {A.}~\bibnamefont {Stegmeir}}, \bibinfo
  {author} {\bibfnamefont {K.}~\bibnamefont {Eder}},\ and\ \bibinfo {author}
  {\bibfnamefont {F.}~\bibnamefont {Jenko}},\ }\href
  {https://doi.org/10.1016/j.cpc.2025.109670} {\bibfield  {journal} {\bibinfo
  {journal} {Computer Physics Communications}\ ,\ \bibinfo {pages} {109670}}
  (\bibinfo {year} {2025})}\BibitemShut {NoStop}%
\bibitem [{\citenamefont {Zholobenko}\ \emph {et~al.}(2024)\citenamefont
  {Zholobenko}, \citenamefont {Zhang}, \citenamefont {Stegmeir}, \citenamefont
  {Pfennig}, \citenamefont {Eder}, \citenamefont {Pitzal}, \citenamefont
  {Ulbl}, \citenamefont {Griener}, \citenamefont {Radovanovic}, \citenamefont
  {Plank} \emph {et~al.}}]{zholobenko2024tokamak}%
  \BibitemOpen
  \bibfield  {author} {\bibinfo {author} {\bibfnamefont {W.}~\bibnamefont
  {Zholobenko}}, \bibinfo {author} {\bibfnamefont {K.}~\bibnamefont {Zhang}},
  \bibinfo {author} {\bibfnamefont {A.}~\bibnamefont {Stegmeir}}, \bibinfo
  {author} {\bibfnamefont {J.}~\bibnamefont {Pfennig}}, \bibinfo {author}
  {\bibfnamefont {K.}~\bibnamefont {Eder}}, \bibinfo {author} {\bibfnamefont
  {C.}~\bibnamefont {Pitzal}}, \bibinfo {author} {\bibfnamefont
  {P.}~\bibnamefont {Ulbl}}, \bibinfo {author} {\bibfnamefont {M.}~\bibnamefont
  {Griener}}, \bibinfo {author} {\bibfnamefont {L.}~\bibnamefont
  {Radovanovic}}, \bibinfo {author} {\bibfnamefont {U.}~\bibnamefont {Plank}},
  \emph {et~al.},\ }\href {https://doi.org/10.1088/1741-4326/ad7611} {\bibfield
   {journal} {\bibinfo  {journal} {Nuclear Fusion}\ }\textbf {\bibinfo {volume}
  {64}},\ \bibinfo {pages} {106066} (\bibinfo {year} {2024})}\BibitemShut
  {NoStop}%
\bibitem [{\citenamefont {Zholobenko}\ \emph {et~al.}(2026)\citenamefont
  {Zholobenko}, \citenamefont {Jenko}, \citenamefont {Zhang}, \citenamefont
  {Ulbl}, \citenamefont {Eder}, \citenamefont {Stegmeir}, \citenamefont
  {Angioni},\ and\ \citenamefont {Manz}}]{zholobenkoFast}%
  \BibitemOpen
  \bibfield  {author} {\bibinfo {author} {\bibfnamefont {W.}~\bibnamefont
  {Zholobenko}}, \bibinfo {author} {\bibfnamefont {F.}~\bibnamefont {Jenko}},
  \bibinfo {author} {\bibfnamefont {K.}~\bibnamefont {Zhang}}, \bibinfo
  {author} {\bibfnamefont {P.}~\bibnamefont {Ulbl}}, \bibinfo {author}
  {\bibfnamefont {K.}~\bibnamefont {Eder}}, \bibinfo {author} {\bibfnamefont
  {A.}~\bibnamefont {Stegmeir}}, \bibinfo {author} {\bibfnamefont
  {C.}~\bibnamefont {Angioni}},\ and\ \bibinfo {author} {\bibfnamefont
  {P.}~\bibnamefont {Manz}},\ }\href {https://doi.org/10.1103/b2s6-b5c1}
  {\bibfield  {journal} {\bibinfo  {journal} {Phys. Rev. Lett.}\ ,\ } (\bibinfo
  {year} {2026})}\BibitemShut {NoStop}%
\bibitem [{\citenamefont {Stegmeir}\ \emph {et~al.}(2016)\citenamefont
  {Stegmeir}, \citenamefont {Coster}, \citenamefont {Maj}, \citenamefont
  {Hallatschek},\ and\ \citenamefont {Lackner}}]{stegmeirFieldLineMap2016}%
  \BibitemOpen
  \bibfield  {author} {\bibinfo {author} {\bibfnamefont {A.}~\bibnamefont
  {Stegmeir}}, \bibinfo {author} {\bibfnamefont {D.}~\bibnamefont {Coster}},
  \bibinfo {author} {\bibfnamefont {O.}~\bibnamefont {Maj}}, \bibinfo {author}
  {\bibfnamefont {K.}~\bibnamefont {Hallatschek}},\ and\ \bibinfo {author}
  {\bibfnamefont {K.}~\bibnamefont {Lackner}},\ }\href
  {https://doi.org/10.1016/j.cpc.2015.09.016} {\bibfield  {journal} {\bibinfo
  {journal} {Computer Physics Communications}\ }\textbf {\bibinfo {volume}
  {198}},\ \bibinfo {pages} {139} (\bibinfo {year} {2016})}\BibitemShut
  {NoStop}%
\bibitem [{\citenamefont {Stegmeir}\ \emph {et~al.}(2017)\citenamefont
  {Stegmeir}, \citenamefont {Maj}, \citenamefont {Coster}, \citenamefont
  {Lackner}, \citenamefont {Held},\ and\ \citenamefont
  {Wiesenberger}}]{stegmeirAdvancesFluxcoordinateIndependent2017}%
  \BibitemOpen
  \bibfield  {author} {\bibinfo {author} {\bibfnamefont {A.}~\bibnamefont
  {Stegmeir}}, \bibinfo {author} {\bibfnamefont {O.}~\bibnamefont {Maj}},
  \bibinfo {author} {\bibfnamefont {D.}~\bibnamefont {Coster}}, \bibinfo
  {author} {\bibfnamefont {K.}~\bibnamefont {Lackner}}, \bibinfo {author}
  {\bibfnamefont {M.}~\bibnamefont {Held}},\ and\ \bibinfo {author}
  {\bibfnamefont {M.}~\bibnamefont {Wiesenberger}},\ }\href
  {https://doi.org/10.1016/j.cpc.2016.12.014} {\bibfield  {journal} {\bibinfo
  {journal} {Computer Physics Communications}\ }\textbf {\bibinfo {volume}
  {213}},\ \bibinfo {pages} {111} (\bibinfo {year} {2017})}\BibitemShut
  {NoStop}%
\bibitem [{\citenamefont {Stegmeir}\ \emph {et~al.}(2023)\citenamefont
  {Stegmeir}, \citenamefont {Body},\ and\ \citenamefont
  {Zholobenko}}]{stegmeirAnalysisLocallyalignedNonaligned2023}%
  \BibitemOpen
  \bibfield  {author} {\bibinfo {author} {\bibfnamefont {A.}~\bibnamefont
  {Stegmeir}}, \bibinfo {author} {\bibfnamefont {T.}~\bibnamefont {Body}},\
  and\ \bibinfo {author} {\bibfnamefont {W.}~\bibnamefont {Zholobenko}},\
  }\href {https://doi.org/10.1016/j.cpc.2023.108801} {\bibfield  {journal}
  {\bibinfo  {journal} {Computer Physics Communications}\ }\textbf {\bibinfo
  {volume} {290}},\ \bibinfo {pages} {108801} (\bibinfo {year}
  {2023})}\BibitemShut {NoStop}%
\bibitem [{\citenamefont {Griener}\ \emph {et~al.}(2020)\citenamefont
  {Griener}, \citenamefont {Wolfrum}, \citenamefont {Birkenmeier},
  \citenamefont {Faitsch}, \citenamefont {Fischer}, \citenamefont {Fuchert},
  \citenamefont {Gil}, \citenamefont {Harrer}, \citenamefont {Manz},
  \citenamefont {Wendler},\ and\ \citenamefont
  {Stroth}}]{grienerContinuousObservationFilaments2020}%
  \BibitemOpen
  \bibfield  {author} {\bibinfo {author} {\bibfnamefont {M.}~\bibnamefont
  {Griener}}, \bibinfo {author} {\bibfnamefont {E.}~\bibnamefont {Wolfrum}},
  \bibinfo {author} {\bibfnamefont {G.}~\bibnamefont {Birkenmeier}}, \bibinfo
  {author} {\bibfnamefont {M.}~\bibnamefont {Faitsch}}, \bibinfo {author}
  {\bibfnamefont {R.}~\bibnamefont {Fischer}}, \bibinfo {author} {\bibfnamefont
  {G.}~\bibnamefont {Fuchert}}, \bibinfo {author} {\bibfnamefont
  {L.}~\bibnamefont {Gil}}, \bibinfo {author} {\bibfnamefont {G.}~\bibnamefont
  {Harrer}}, \bibinfo {author} {\bibfnamefont {P.}~\bibnamefont {Manz}},
  \bibinfo {author} {\bibfnamefont {D.}~\bibnamefont {Wendler}},\ and\ \bibinfo
  {author} {\bibfnamefont {U.}~\bibnamefont {Stroth}},\ }\href
  {https://doi.org/10.1016/j.nme.2020.100854} {\bibfield  {journal} {\bibinfo
  {journal} {Nuclear Materials and Energy}\ }\textbf {\bibinfo {volume} {25}},\
  \bibinfo {pages} {100854} (\bibinfo {year} {2020})}\BibitemShut {NoStop}%
\bibitem [{\citenamefont {Fischer}\ \emph {et~al.}(2010)\citenamefont
  {Fischer}, \citenamefont {Fuchs}, \citenamefont {Kurzan}, \citenamefont
  {Suttrop}, \citenamefont {Wolfrum},\ and\ \citenamefont
  {Team}}]{fischer2010integrated}%
  \BibitemOpen
  \bibfield  {author} {\bibinfo {author} {\bibfnamefont {R.}~\bibnamefont
  {Fischer}}, \bibinfo {author} {\bibfnamefont {C.}~\bibnamefont {Fuchs}},
  \bibinfo {author} {\bibfnamefont {B.}~\bibnamefont {Kurzan}}, \bibinfo
  {author} {\bibfnamefont {W.}~\bibnamefont {Suttrop}}, \bibinfo {author}
  {\bibfnamefont {E.}~\bibnamefont {Wolfrum}},\ and\ \bibinfo {author}
  {\bibfnamefont {A.~U.}\ \bibnamefont {Team}},\ }\href
  {https://doi.org/10.13182/FST10-110} {\bibfield  {journal} {\bibinfo
  {journal} {Fusion science and technology}\ }\textbf {\bibinfo {volume}
  {58}},\ \bibinfo {pages} {675} (\bibinfo {year} {2010})}\BibitemShut
  {NoStop}%
\bibitem [{\citenamefont {Eder}\ \emph {et~al.}(2025)\citenamefont {Eder},
  \citenamefont {Stegmeir}, \citenamefont {Zholobenko}, \citenamefont
  {Pfennig}, \citenamefont {Brida}, \citenamefont {Grenfell}, \citenamefont
  {Jenko},\ and\ \citenamefont {{ASDEX Upgrade Team}}}]{eder2025self}%
  \BibitemOpen
  \bibfield  {author} {\bibinfo {author} {\bibfnamefont {K.}~\bibnamefont
  {Eder}}, \bibinfo {author} {\bibfnamefont {A.}~\bibnamefont {Stegmeir}},
  \bibinfo {author} {\bibfnamefont {W.}~\bibnamefont {Zholobenko}}, \bibinfo
  {author} {\bibfnamefont {J.}~\bibnamefont {Pfennig}}, \bibinfo {author}
  {\bibfnamefont {D.}~\bibnamefont {Brida}}, \bibinfo {author} {\bibfnamefont
  {G.}~\bibnamefont {Grenfell}}, \bibinfo {author} {\bibfnamefont
  {F.}~\bibnamefont {Jenko}},\ and\ \bibinfo {author} {\bibnamefont {{ASDEX
  Upgrade Team}}},\ }\href {https://doi.org/10.1088/1361-6587/add8ba}
  {\bibfield  {journal} {\bibinfo  {journal} {Plasma Physics and Controlled
  Fusion}\ }\textbf {\bibinfo {volume} {67}},\ \bibinfo {pages} {065034}
  (\bibinfo {year} {2025})}\BibitemShut {NoStop}%
\bibitem [{\citenamefont {Hirshman}\ and\ \citenamefont
  {Sigmar}(1981)}]{Hirshman1981}%
  \BibitemOpen
  \bibfield  {author} {\bibinfo {author} {\bibfnamefont {S.}~\bibnamefont
  {Hirshman}}\ and\ \bibinfo {author} {\bibfnamefont {D.}~\bibnamefont
  {Sigmar}},\ }\href {https://doi.org/10.1088/0029-5515/21/9/003} {\bibfield
  {journal} {\bibinfo  {journal} {Nuclear Fusion}\ }\textbf {\bibinfo {volume}
  {21}},\ \bibinfo {pages} {1079} (\bibinfo {year} {1981})}\BibitemShut
  {NoStop}%
\bibitem [{\citenamefont {Rozhansky}\ \emph {et~al.}(2009)\citenamefont
  {Rozhansky}, \citenamefont {Kaveeva}, \citenamefont {Molchanov},
  \citenamefont {Veselova}, \citenamefont {Voskoboynikov}, \citenamefont
  {Coster}, \citenamefont {Counsell}, \citenamefont {Kirk}, \citenamefont
  {Lisgo},\ and\ \citenamefont {and}}]{Rozhansky2009}%
  \BibitemOpen
  \bibfield  {author} {\bibinfo {author} {\bibfnamefont {V.}~\bibnamefont
  {Rozhansky}}, \bibinfo {author} {\bibfnamefont {E.}~\bibnamefont {Kaveeva}},
  \bibinfo {author} {\bibfnamefont {P.}~\bibnamefont {Molchanov}}, \bibinfo
  {author} {\bibfnamefont {I.}~\bibnamefont {Veselova}}, \bibinfo {author}
  {\bibfnamefont {S.}~\bibnamefont {Voskoboynikov}}, \bibinfo {author}
  {\bibfnamefont {D.}~\bibnamefont {Coster}}, \bibinfo {author} {\bibfnamefont
  {G.}~\bibnamefont {Counsell}}, \bibinfo {author} {\bibfnamefont
  {A.}~\bibnamefont {Kirk}}, \bibinfo {author} {\bibfnamefont {S.}~\bibnamefont
  {Lisgo}},\ and\ \bibinfo {author} {\bibnamefont {and}},\ }\href
  {https://doi.org/10.1088/0029-5515/49/2/025007} {\bibfield  {journal}
  {\bibinfo  {journal} {Nuclear Fusion}\ }\textbf {\bibinfo {volume} {49}},\
  \bibinfo {pages} {025007} (\bibinfo {year} {2009})}\BibitemShut {NoStop}%
\bibitem [{\citenamefont {Hammett}\ and\ \citenamefont
  {Perkins}(1990)}]{hammett1990fluid}%
  \BibitemOpen
  \bibfield  {author} {\bibinfo {author} {\bibfnamefont {G.~W.}\ \bibnamefont
  {Hammett}}\ and\ \bibinfo {author} {\bibfnamefont {F.~W.}\ \bibnamefont
  {Perkins}},\ }\href {https://doi.org/10.1103/PhysRevLett.64.3019} {\bibfield
  {journal} {\bibinfo  {journal} {Physical review letters}\ }\textbf {\bibinfo
  {volume} {64}},\ \bibinfo {pages} {3019} (\bibinfo {year}
  {1990})}\BibitemShut {NoStop}%
\bibitem [{\citenamefont {Pitzal}\ \emph {et~al.}(2023)\citenamefont {Pitzal},
  \citenamefont {Stegmeir}, \citenamefont {Zholobenko}, \citenamefont {Zhang},\
  and\ \citenamefont {Jenko}}]{pitzal2023landau}%
  \BibitemOpen
  \bibfield  {author} {\bibinfo {author} {\bibfnamefont {C.}~\bibnamefont
  {Pitzal}}, \bibinfo {author} {\bibfnamefont {A.}~\bibnamefont {Stegmeir}},
  \bibinfo {author} {\bibfnamefont {W.}~\bibnamefont {Zholobenko}}, \bibinfo
  {author} {\bibfnamefont {K.}~\bibnamefont {Zhang}},\ and\ \bibinfo {author}
  {\bibfnamefont {F.}~\bibnamefont {Jenko}},\ }\bibfield  {journal} {\bibinfo
  {journal} {Physics of Plasmas}\ }\textbf {\bibinfo {volume} {30}},\ \href
  {https://doi.org/10.1063/5.0171184} {10.1063/5.0171184} (\bibinfo {year}
  {2023})\BibitemShut {NoStop}%
\bibitem [{\citenamefont {Braginskii}(1965)}]{braginskii1965transport}%
  \BibitemOpen
  \bibfield  {author} {\bibinfo {author} {\bibfnamefont {S.}~\bibnamefont
  {Braginskii}},\ }\href@noop {} {\bibfield  {journal} {\bibinfo  {journal}
  {Reviews of Plasma Physics}\ }\textbf {\bibinfo {volume} {1}},\ \bibinfo
  {pages} {205} (\bibinfo {year} {1965})}\BibitemShut {NoStop}%
\bibitem [{\citenamefont {Harrer}\ \emph {et~al.}(2022)\citenamefont {Harrer},
  \citenamefont {Faitsch}, \citenamefont {Radovanovic}, \citenamefont
  {Wolfrum}, \citenamefont {Albert}, \citenamefont {Cathey}, \citenamefont
  {Cavedon}, \citenamefont {Dunne}, \citenamefont {Eich}, \citenamefont
  {Fischer}, \citenamefont {Griener}, \citenamefont {Hoelzl}, \citenamefont
  {Labit}, \citenamefont {Meyer}, \citenamefont {Aumayr}, \citenamefont {{The
  Asdex Upgrade Team}},\ and\ \citenamefont {{The EUROfusion MST1
  Team}}}]{harrerQuasicontinuousExhaustScenario2022}%
  \BibitemOpen
  \bibfield  {author} {\bibinfo {author} {\bibfnamefont {G.~F.}\ \bibnamefont
  {Harrer}}, \bibinfo {author} {\bibfnamefont {M.}~\bibnamefont {Faitsch}},
  \bibinfo {author} {\bibfnamefont {L.}~\bibnamefont {Radovanovic}}, \bibinfo
  {author} {\bibfnamefont {E.}~\bibnamefont {Wolfrum}}, \bibinfo {author}
  {\bibfnamefont {C.}~\bibnamefont {Albert}}, \bibinfo {author} {\bibfnamefont
  {A.}~\bibnamefont {Cathey}}, \bibinfo {author} {\bibfnamefont
  {M.}~\bibnamefont {Cavedon}}, \bibinfo {author} {\bibfnamefont
  {M.}~\bibnamefont {Dunne}}, \bibinfo {author} {\bibfnamefont
  {T.}~\bibnamefont {Eich}}, \bibinfo {author} {\bibfnamefont {R.}~\bibnamefont
  {Fischer}}, \bibinfo {author} {\bibfnamefont {M.}~\bibnamefont {Griener}},
  \bibinfo {author} {\bibfnamefont {M.}~\bibnamefont {Hoelzl}}, \bibinfo
  {author} {\bibfnamefont {B.}~\bibnamefont {Labit}}, \bibinfo {author}
  {\bibfnamefont {H.}~\bibnamefont {Meyer}}, \bibinfo {author} {\bibfnamefont
  {F.}~\bibnamefont {Aumayr}}, \bibinfo {author} {\bibnamefont {{The Asdex
  Upgrade Team}}},\ and\ \bibinfo {author} {\bibnamefont {{The EUROfusion MST1
  Team}}},\ }\href {https://doi.org/10.1103/PhysRevLett.129.165001} {\bibfield
  {journal} {\bibinfo  {journal} {Physical Review Letters}\ }\textbf {\bibinfo
  {volume} {129}},\ \bibinfo {pages} {165001} (\bibinfo {year}
  {2022})}\BibitemShut {NoStop}%
\bibitem [{\citenamefont {Sun}\ \emph {et~al.}(2015)\citenamefont {Sun},
  \citenamefont {Wolfrum}, \citenamefont {Eich}, \citenamefont {Kurzan},
  \citenamefont {Potzel}, \citenamefont {Stroth}, \citenamefont {Team} \emph
  {et~al.}}]{sun2015study}%
  \BibitemOpen
  \bibfield  {author} {\bibinfo {author} {\bibfnamefont {H.}~\bibnamefont
  {Sun}}, \bibinfo {author} {\bibfnamefont {E.}~\bibnamefont {Wolfrum}},
  \bibinfo {author} {\bibfnamefont {T.}~\bibnamefont {Eich}}, \bibinfo {author}
  {\bibfnamefont {B.}~\bibnamefont {Kurzan}}, \bibinfo {author} {\bibfnamefont
  {S.}~\bibnamefont {Potzel}}, \bibinfo {author} {\bibfnamefont
  {U.}~\bibnamefont {Stroth}}, \bibinfo {author} {\bibfnamefont {A.~U.}\
  \bibnamefont {Team}}, \emph {et~al.},\ }\href
  {https://doi.org/10.1088/0741-3335/57/12/125011} {\bibfield  {journal}
  {\bibinfo  {journal} {Plasma Physics and Controlled Fusion}\ }\textbf
  {\bibinfo {volume} {57}},\ \bibinfo {pages} {125011} (\bibinfo {year}
  {2015})}\BibitemShut {NoStop}%
\bibitem [{\citenamefont {Manz}\ \emph {et~al.}(2015)\citenamefont {Manz},
  \citenamefont {Ribeiro}, \citenamefont {Scott}, \citenamefont {Birkenmeier},
  \citenamefont {Carralero}, \citenamefont {Fuchert}, \citenamefont
  {M{\"u}ller}, \citenamefont {M{\"u}ller}, \citenamefont {Stroth},\ and\
  \citenamefont {Wolfrum}}]{manz2015origin}%
  \BibitemOpen
  \bibfield  {author} {\bibinfo {author} {\bibfnamefont {P.}~\bibnamefont
  {Manz}}, \bibinfo {author} {\bibfnamefont {T.}~\bibnamefont {Ribeiro}},
  \bibinfo {author} {\bibfnamefont {B.}~\bibnamefont {Scott}}, \bibinfo
  {author} {\bibfnamefont {G.}~\bibnamefont {Birkenmeier}}, \bibinfo {author}
  {\bibfnamefont {D.}~\bibnamefont {Carralero}}, \bibinfo {author}
  {\bibfnamefont {G.}~\bibnamefont {Fuchert}}, \bibinfo {author} {\bibfnamefont
  {S.}~\bibnamefont {M{\"u}ller}}, \bibinfo {author} {\bibfnamefont
  {H.}~\bibnamefont {M{\"u}ller}}, \bibinfo {author} {\bibfnamefont
  {U.}~\bibnamefont {Stroth}},\ and\ \bibinfo {author} {\bibfnamefont
  {E.}~\bibnamefont {Wolfrum}},\ }\href {https://doi.org/10.1063/1.4908272}
  {\bibfield  {journal} {\bibinfo  {journal} {Physics of Plasmas}\ }\textbf
  {\bibinfo {volume} {22}},\ \bibinfo {pages} {022308} (\bibinfo {year}
  {2015})}\BibitemShut {NoStop}%
\bibitem [{\citenamefont {Aleynikova}\ and\ \citenamefont
  {Zocco}(2017)}]{aleynikova2017quantitative}%
  \BibitemOpen
  \bibfield  {author} {\bibinfo {author} {\bibfnamefont {K.}~\bibnamefont
  {Aleynikova}}\ and\ \bibinfo {author} {\bibfnamefont {A.}~\bibnamefont
  {Zocco}},\ }\bibfield  {journal} {\bibinfo  {journal} {Physics of Plasmas}\
  }\textbf {\bibinfo {volume} {24}},\ \href {https://doi.org/10.1063/1.5000052}
  {10.1063/1.5000052} (\bibinfo {year} {2017})\BibitemShut {NoStop}%
\bibitem [{\citenamefont {Ara}\ \emph {et~al.}(1978)\citenamefont {Ara},
  \citenamefont {Basu}, \citenamefont {Coppi}, \citenamefont {Laval},
  \citenamefont {Rosenbluth},\ and\ \citenamefont {Waddell}}]{ara1978magnetic}%
  \BibitemOpen
  \bibfield  {author} {\bibinfo {author} {\bibfnamefont {G.}~\bibnamefont
  {Ara}}, \bibinfo {author} {\bibfnamefont {B.}~\bibnamefont {Basu}}, \bibinfo
  {author} {\bibfnamefont {B.}~\bibnamefont {Coppi}}, \bibinfo {author}
  {\bibfnamefont {G.}~\bibnamefont {Laval}}, \bibinfo {author} {\bibfnamefont
  {M.}~\bibnamefont {Rosenbluth}},\ and\ \bibinfo {author} {\bibfnamefont
  {B.}~\bibnamefont {Waddell}},\ }\href
  {https://doi.org/10.1016/S0003-4916(78)80007-4} {\bibfield  {journal}
  {\bibinfo  {journal} {Annals of Physics}\ }\textbf {\bibinfo {volume}
  {112}},\ \bibinfo {pages} {443} (\bibinfo {year} {1978})}\BibitemShut
  {NoStop}%
\bibitem [{\citenamefont {Porcelli}\ \emph {et~al.}(1986)\citenamefont
  {Porcelli}, \citenamefont {Migliuolo} \emph {et~al.}}]{porcelli1986ion}%
  \BibitemOpen
  \bibfield  {author} {\bibinfo {author} {\bibfnamefont {F.}~\bibnamefont
  {Porcelli}}, \bibinfo {author} {\bibfnamefont {S.}~\bibnamefont {Migliuolo}},
  \emph {et~al.},\ }\href {https://doi.org/10.1063/1.865643} {\bibfield
  {journal} {\bibinfo  {journal} {The Physics of fluids}\ }\textbf {\bibinfo
  {volume} {29}},\ \bibinfo {pages} {1741} (\bibinfo {year}
  {1986})}\BibitemShut {NoStop}%
\bibitem [{\citenamefont {Kotschenreuther}\ \emph {et~al.}(2019)\citenamefont
  {Kotschenreuther}, \citenamefont {Liu}, \citenamefont {Hatch}, \citenamefont
  {Mahajan}, \citenamefont {Zheng}, \citenamefont {Diallo}, \citenamefont
  {Groebner}, \citenamefont {Hillesheim}, \citenamefont {Maggi}, \citenamefont
  {Giroud} \emph {et~al.}}]{kotschenreuther2019gyrokinetic}%
  \BibitemOpen
  \bibfield  {author} {\bibinfo {author} {\bibfnamefont {M.}~\bibnamefont
  {Kotschenreuther}}, \bibinfo {author} {\bibfnamefont {X.}~\bibnamefont
  {Liu}}, \bibinfo {author} {\bibfnamefont {D.}~\bibnamefont {Hatch}}, \bibinfo
  {author} {\bibfnamefont {S.}~\bibnamefont {Mahajan}}, \bibinfo {author}
  {\bibfnamefont {L.}~\bibnamefont {Zheng}}, \bibinfo {author} {\bibfnamefont
  {A.}~\bibnamefont {Diallo}}, \bibinfo {author} {\bibfnamefont
  {R.}~\bibnamefont {Groebner}}, \bibinfo {author} {\bibfnamefont
  {J.}~\bibnamefont {Hillesheim}}, \bibinfo {author} {\bibfnamefont
  {C.}~\bibnamefont {Maggi}}, \bibinfo {author} {\bibfnamefont
  {C.}~\bibnamefont {Giroud}}, \emph {et~al.},\ }\href
  {https://doi.org/10.1088/1741-4326/ab1fa2} {\bibfield  {journal} {\bibinfo
  {journal} {Nuclear Fusion}\ }\textbf {\bibinfo {volume} {59}},\ \bibinfo
  {pages} {096001} (\bibinfo {year} {2019})}\BibitemShut {NoStop}%
\bibitem [{\citenamefont {Gupta}\ \emph {et~al.}(2002)\citenamefont {Gupta},
  \citenamefont {Callen},\ and\ \citenamefont {Hegna}}]{gupta2002violating}%
  \BibitemOpen
  \bibfield  {author} {\bibinfo {author} {\bibfnamefont {S.}~\bibnamefont
  {Gupta}}, \bibinfo {author} {\bibfnamefont {J.~D.}\ \bibnamefont {Callen}},\
  and\ \bibinfo {author} {\bibfnamefont {C.~C.}\ \bibnamefont {Hegna}},\ }\href
  {https://doi.org/10.1063/1.1487865} {\bibfield  {journal} {\bibinfo
  {journal} {Physics of Plasmas}\ }\textbf {\bibinfo {volume} {9}},\ \bibinfo
  {pages} {3395} (\bibinfo {year} {2002})}\BibitemShut {NoStop}%
\bibitem [{\citenamefont {Myra}\ and\ \citenamefont
  {D’Ippolito}(2005)}]{myra2005edge}%
  \BibitemOpen
  \bibfield  {author} {\bibinfo {author} {\bibfnamefont {J.}~\bibnamefont
  {Myra}}\ and\ \bibinfo {author} {\bibfnamefont {D.}~\bibnamefont
  {D’Ippolito}},\ }\bibfield  {journal} {\bibinfo  {journal} {Physics of
  Plasmas}\ }\textbf {\bibinfo {volume} {12}},\ \href
  {https://doi.org/10.1063/1.2048847} {10.1063/1.2048847} (\bibinfo {year}
  {2005})\BibitemShut {NoStop}%
\bibitem [{\citenamefont {Doerk}\ \emph {et~al.}(2016)\citenamefont {Doerk},
  \citenamefont {Challis}, \citenamefont {Citrin}, \citenamefont {Garcia},
  \citenamefont {G{\"o}rler}, \citenamefont {Jenko},\ and\ \citenamefont
  {Contributors}}]{doerk2016gyrokinetic}%
  \BibitemOpen
  \bibfield  {author} {\bibinfo {author} {\bibfnamefont {H.}~\bibnamefont
  {Doerk}}, \bibinfo {author} {\bibfnamefont {C.}~\bibnamefont {Challis}},
  \bibinfo {author} {\bibfnamefont {J.}~\bibnamefont {Citrin}}, \bibinfo
  {author} {\bibfnamefont {J.}~\bibnamefont {Garcia}}, \bibinfo {author}
  {\bibfnamefont {T.}~\bibnamefont {G{\"o}rler}}, \bibinfo {author}
  {\bibfnamefont {F.}~\bibnamefont {Jenko}},\ and\ \bibinfo {author}
  {\bibfnamefont {J.}~\bibnamefont {Contributors}},\ }\href
  {https://doi.org/10.1088/0741-3335/58/11/115005} {\bibfield  {journal}
  {\bibinfo  {journal} {Plasma Physics and Controlled Fusion}\ }\textbf
  {\bibinfo {volume} {58}},\ \bibinfo {pages} {115005} (\bibinfo {year}
  {2016})}\BibitemShut {NoStop}%
\bibitem [{\citenamefont {Myra}\ \emph {et~al.}(2000)\citenamefont {Myra},
  \citenamefont {D'Ippolito}, \citenamefont {Xu},\ and\ \citenamefont
  {Cohen}}]{MyraDippolito2000}%
  \BibitemOpen
  \bibfield  {author} {\bibinfo {author} {\bibfnamefont {J.~R.}\ \bibnamefont
  {Myra}}, \bibinfo {author} {\bibfnamefont {D.~A.}\ \bibnamefont
  {D'Ippolito}}, \bibinfo {author} {\bibfnamefont {X.~Q.}\ \bibnamefont {Xu}},\
  and\ \bibinfo {author} {\bibfnamefont {R.~H.}\ \bibnamefont {Cohen}},\ }\href
  {https://doi.org/10.1063/1.874125} {\bibfield  {journal} {\bibinfo  {journal}
  {Physics of Plasmas}\ }\textbf {\bibinfo {volume} {7}},\ \bibinfo {pages}
  {2290} (\bibinfo {year} {2000})}\BibitemShut {NoStop}%
\bibitem [{\citenamefont {Krasheninnikov}(2001)}]{krasheninnikov2001scrape}%
  \BibitemOpen
  \bibfield  {author} {\bibinfo {author} {\bibfnamefont {S.~I.}\ \bibnamefont
  {Krasheninnikov}},\ }\href {https://doi.org/10.1016/S0375-9601(01)00252-3}
  {\bibfield  {journal} {\bibinfo  {journal} {Physics Letters A}\ }\textbf
  {\bibinfo {volume} {283}},\ \bibinfo {pages} {368} (\bibinfo {year}
  {2001})}\BibitemShut {NoStop}%
\bibitem [{\citenamefont {Radovanovic}\ \emph {et~al.}(2022)\citenamefont
  {Radovanovic}, \citenamefont {Dunne}, \citenamefont {Wolfrum}, \citenamefont
  {Harrer}, \citenamefont {Faitsch}, \citenamefont {Fischer},\ and\
  \citenamefont {Aumayr}}]{radovanovicDevelopingPhysicsUnderstanding2022}%
  \BibitemOpen
  \bibfield  {author} {\bibinfo {author} {\bibfnamefont {L.}~\bibnamefont
  {Radovanovic}}, \bibinfo {author} {\bibfnamefont {M.}~\bibnamefont {Dunne}},
  \bibinfo {author} {\bibfnamefont {E.}~\bibnamefont {Wolfrum}}, \bibinfo
  {author} {\bibfnamefont {G.}~\bibnamefont {Harrer}}, \bibinfo {author}
  {\bibfnamefont {M.}~\bibnamefont {Faitsch}}, \bibinfo {author} {\bibfnamefont
  {R.}~\bibnamefont {Fischer}},\ and\ \bibinfo {author} {\bibfnamefont
  {F.}~\bibnamefont {Aumayr}},\ }\bibfield  {journal} {\bibinfo  {journal}
  {Nuclear Fusion}\ }\href {https://doi.org/10.1088/1741-4326/ac6d6a}
  {10.1088/1741-4326/ac6d6a} (\bibinfo {year} {2022})\BibitemShut {NoStop}%
\bibitem [{\citenamefont {Gil}\ \emph {et~al.}(2020)\citenamefont {Gil},
  \citenamefont {Silva}, \citenamefont {Happel}, \citenamefont {Birkenmeier},
  \citenamefont {Conway}, \citenamefont {Guimarais}, \citenamefont
  {Kallenbach}, \citenamefont {P{\"u}tterich}, \citenamefont {Santos},
  \citenamefont {Schneider} \emph {et~al.}}]{gil2020stationary}%
  \BibitemOpen
  \bibfield  {author} {\bibinfo {author} {\bibfnamefont {L.}~\bibnamefont
  {Gil}}, \bibinfo {author} {\bibfnamefont {C.}~\bibnamefont {Silva}}, \bibinfo
  {author} {\bibfnamefont {T.}~\bibnamefont {Happel}}, \bibinfo {author}
  {\bibfnamefont {G.}~\bibnamefont {Birkenmeier}}, \bibinfo {author}
  {\bibfnamefont {G.~D.}\ \bibnamefont {Conway}}, \bibinfo {author}
  {\bibfnamefont {L.}~\bibnamefont {Guimarais}}, \bibinfo {author}
  {\bibfnamefont {A.}~\bibnamefont {Kallenbach}}, \bibinfo {author}
  {\bibfnamefont {T.}~\bibnamefont {P{\"u}tterich}}, \bibinfo {author}
  {\bibfnamefont {J.}~\bibnamefont {Santos}}, \bibinfo {author} {\bibfnamefont
  {P.~A.}\ \bibnamefont {Schneider}}, \emph {et~al.},\ }\href
  {https://doi.org/10.1088/1741-4326/ab7d1b} {\bibfield  {journal} {\bibinfo
  {journal} {Nuclear Fusion}\ }\textbf {\bibinfo {volume} {60}},\ \bibinfo
  {pages} {054003} (\bibinfo {year} {2020})}\BibitemShut {NoStop}%
\bibitem [{\citenamefont {Snyder}\ \emph {et~al.}(2011)\citenamefont {Snyder},
  \citenamefont {Groebner}, \citenamefont {Hughes}, \citenamefont {Osborne},
  \citenamefont {Beurskens}, \citenamefont {Leonard}, \citenamefont {Wilson},\
  and\ \citenamefont {Xu}}]{snyder2011first}%
  \BibitemOpen
  \bibfield  {author} {\bibinfo {author} {\bibfnamefont {P.}~\bibnamefont
  {Snyder}}, \bibinfo {author} {\bibfnamefont {R.}~\bibnamefont {Groebner}},
  \bibinfo {author} {\bibfnamefont {J.}~\bibnamefont {Hughes}}, \bibinfo
  {author} {\bibfnamefont {T.}~\bibnamefont {Osborne}}, \bibinfo {author}
  {\bibfnamefont {M.}~\bibnamefont {Beurskens}}, \bibinfo {author}
  {\bibfnamefont {A.}~\bibnamefont {Leonard}}, \bibinfo {author} {\bibfnamefont
  {H.}~\bibnamefont {Wilson}},\ and\ \bibinfo {author} {\bibfnamefont
  {X.}~\bibnamefont {Xu}},\ }\href
  {https://doi.org/10.1088/0029-5515/51/10/103016} {\bibfield  {journal}
  {\bibinfo  {journal} {Nuclear Fusion}\ }\textbf {\bibinfo {volume} {51}},\
  \bibinfo {pages} {103016} (\bibinfo {year} {2011})}\BibitemShut {NoStop}%
\bibitem [{\citenamefont {Dickinson}\ \emph {et~al.}(2012)\citenamefont
  {Dickinson}, \citenamefont {Roach}, \citenamefont {Saarelma}, \citenamefont
  {Scannell}, \citenamefont {Kirk},\ and\ \citenamefont
  {Wilson}}]{Dickinson2012}%
  \BibitemOpen
  \bibfield  {author} {\bibinfo {author} {\bibfnamefont {D.}~\bibnamefont
  {Dickinson}}, \bibinfo {author} {\bibfnamefont {C.~M.}\ \bibnamefont
  {Roach}}, \bibinfo {author} {\bibfnamefont {S.}~\bibnamefont {Saarelma}},
  \bibinfo {author} {\bibfnamefont {R.}~\bibnamefont {Scannell}}, \bibinfo
  {author} {\bibfnamefont {A.}~\bibnamefont {Kirk}},\ and\ \bibinfo {author}
  {\bibfnamefont {H.~R.}\ \bibnamefont {Wilson}},\ }\href
  {https://doi.org/10.1103/physrevlett.108.135002} {\bibfield  {journal}
  {\bibinfo  {journal} {Physical Review Letters}\ }\textbf {\bibinfo {volume}
  {108}},\ \bibinfo {pages} {135002} (\bibinfo {year} {2012})}\BibitemShut
  {NoStop}%
\bibitem [{\citenamefont {Wesson}\ and\ \citenamefont
  {Campbell}(2011)}]{wessonTokamaks2011}%
  \BibitemOpen
  \bibfield  {author} {\bibinfo {author} {\bibfnamefont {J.}~\bibnamefont
  {Wesson}}\ and\ \bibinfo {author} {\bibfnamefont {D.~J.}\ \bibnamefont
  {Campbell}},\ }\href@noop {} {\emph {\bibinfo {title} {Tokamaks}}},\ \bibinfo
  {edition} {4th}\ ed.,\ \bibinfo {series} {Oxford Science Publications}\ No.\
  \bibinfo {number} {149}\ (\bibinfo  {publisher} {Oxford University Press},\
  \bibinfo {year} {2011})\BibitemShut {NoStop}%
\bibitem [{\citenamefont {Callen}\ \emph {et~al.}(2010)\citenamefont {Callen},
  \citenamefont {Groebner}, \citenamefont {Osborne}, \citenamefont {Canik},
  \citenamefont {Owen}, \citenamefont {Pankin}, \citenamefont {Rafiq},
  \citenamefont {Rognlien},\ and\ \citenamefont {Stacey}}]{callen2010analysis}%
  \BibitemOpen
  \bibfield  {author} {\bibinfo {author} {\bibfnamefont {J.}~\bibnamefont
  {Callen}}, \bibinfo {author} {\bibfnamefont {R.}~\bibnamefont {Groebner}},
  \bibinfo {author} {\bibfnamefont {T.}~\bibnamefont {Osborne}}, \bibinfo
  {author} {\bibfnamefont {J.}~\bibnamefont {Canik}}, \bibinfo {author}
  {\bibfnamefont {L.~W.}\ \bibnamefont {Owen}}, \bibinfo {author}
  {\bibfnamefont {A.}~\bibnamefont {Pankin}}, \bibinfo {author} {\bibfnamefont
  {T.}~\bibnamefont {Rafiq}}, \bibinfo {author} {\bibfnamefont
  {T.}~\bibnamefont {Rognlien}},\ and\ \bibinfo {author} {\bibfnamefont
  {W.}~\bibnamefont {Stacey}},\ }\href@noop {} {\bibfield  {journal} {\bibinfo
  {journal} {Nuclear fusion}\ }\textbf {\bibinfo {volume} {50}},\ \bibinfo
  {pages} {064004} (\bibinfo {year} {2010})}\BibitemShut {NoStop}%
\bibitem [{\citenamefont {Hahm}\ and\ \citenamefont
  {Diamond}(2018)}]{hahm2018mesoscopic}%
  \BibitemOpen
  \bibfield  {author} {\bibinfo {author} {\bibfnamefont {T.}~\bibnamefont
  {Hahm}}\ and\ \bibinfo {author} {\bibfnamefont {P.}~\bibnamefont {Diamond}},\
  }\href@noop {} {\bibfield  {journal} {\bibinfo  {journal} {Journal of the
  Korean Physical Society}\ }\textbf {\bibinfo {volume} {73}},\ \bibinfo
  {pages} {747} (\bibinfo {year} {2018})}\BibitemShut {NoStop}%
\bibitem [{\citenamefont {Ashourvan}\ and\ \citenamefont
  {Diamond}(2016)}]{ashourvan2016mesoscopic}%
  \BibitemOpen
  \bibfield  {author} {\bibinfo {author} {\bibfnamefont {A.}~\bibnamefont
  {Ashourvan}}\ and\ \bibinfo {author} {\bibfnamefont {P.}~\bibnamefont
  {Diamond}},\ }\href {https://doi.org/10.1103/PhysRevE.94.051202} {\bibfield
  {journal} {\bibinfo  {journal} {Physical Review E}\ }\textbf {\bibinfo
  {volume} {94}},\ \bibinfo {pages} {051202} (\bibinfo {year}
  {2016})}\BibitemShut {NoStop}%
\bibitem [{\citenamefont {Cathey}\ \emph {et~al.}(2022)\citenamefont {Cathey},
  \citenamefont {Hoelzl}, \citenamefont {Harrer}, \citenamefont {Dunne},
  \citenamefont {Huijsmans}, \citenamefont {Lackner}, \citenamefont {Pamela},
  \citenamefont {Wolfrum}, \citenamefont {G{\"u}nter}, \citenamefont {team}
  \emph {et~al.}}]{cathey2022mhd}%
  \BibitemOpen
  \bibfield  {author} {\bibinfo {author} {\bibfnamefont {A.}~\bibnamefont
  {Cathey}}, \bibinfo {author} {\bibfnamefont {M.}~\bibnamefont {Hoelzl}},
  \bibinfo {author} {\bibfnamefont {G.}~\bibnamefont {Harrer}}, \bibinfo
  {author} {\bibfnamefont {M.}~\bibnamefont {Dunne}}, \bibinfo {author}
  {\bibfnamefont {G.}~\bibnamefont {Huijsmans}}, \bibinfo {author}
  {\bibfnamefont {K.}~\bibnamefont {Lackner}}, \bibinfo {author} {\bibfnamefont
  {S.}~\bibnamefont {Pamela}}, \bibinfo {author} {\bibfnamefont
  {E.}~\bibnamefont {Wolfrum}}, \bibinfo {author} {\bibfnamefont
  {S.}~\bibnamefont {G{\"u}nter}}, \bibinfo {author} {\bibfnamefont
  {J.}~\bibnamefont {team}}, \emph {et~al.},\ }\href
  {https://doi.org/10.1088/1361-6587/ac5b4b} {\bibfield  {journal} {\bibinfo
  {journal} {Plasma Physics and Controlled Fusion}\ }\textbf {\bibinfo {volume}
  {64}},\ \bibinfo {pages} {054011} (\bibinfo {year} {2022})}\BibitemShut
  {NoStop}%
\bibitem [{\citenamefont {Diamond}\ and\ \citenamefont
  {Kim}(1991)}]{diamond1991theory}%
  \BibitemOpen
  \bibfield  {author} {\bibinfo {author} {\bibfnamefont {P.}~\bibnamefont
  {Diamond}}\ and\ \bibinfo {author} {\bibfnamefont {Y.-B.}\ \bibnamefont
  {Kim}},\ }\href {https://doi.org/10.1063/1.859681} {\bibfield  {journal}
  {\bibinfo  {journal} {Physics of Fluids B: Plasma Physics}\ }\textbf
  {\bibinfo {volume} {3}},\ \bibinfo {pages} {1626} (\bibinfo {year}
  {1991})}\BibitemShut {NoStop}%
\bibitem [{\citenamefont {Miller}\ \emph {et~al.}(2025)\citenamefont {Miller},
  \citenamefont {Hughes}, \citenamefont {Eich}, \citenamefont {Tynan},
  \citenamefont {Manz}, \citenamefont {Body}, \citenamefont {Silvagni},
  \citenamefont {Grover}, \citenamefont {Hubbard}, \citenamefont {Cavallaro}
  \emph {et~al.}}]{miller2025determination}%
  \BibitemOpen
  \bibfield  {author} {\bibinfo {author} {\bibfnamefont {M.~A.}\ \bibnamefont
  {Miller}}, \bibinfo {author} {\bibfnamefont {J.~W.}\ \bibnamefont {Hughes}},
  \bibinfo {author} {\bibfnamefont {T.}~\bibnamefont {Eich}}, \bibinfo {author}
  {\bibfnamefont {G.~R.}\ \bibnamefont {Tynan}}, \bibinfo {author}
  {\bibfnamefont {P.}~\bibnamefont {Manz}}, \bibinfo {author} {\bibfnamefont
  {T.}~\bibnamefont {Body}}, \bibinfo {author} {\bibfnamefont {D.}~\bibnamefont
  {Silvagni}}, \bibinfo {author} {\bibfnamefont {O.}~\bibnamefont {Grover}},
  \bibinfo {author} {\bibfnamefont {A.~E.}\ \bibnamefont {Hubbard}}, \bibinfo
  {author} {\bibfnamefont {A.}~\bibnamefont {Cavallaro}}, \emph {et~al.},\
  }\href {https://doi.org/10.1088/1741-4326/adc9c3} {\bibfield  {journal}
  {\bibinfo  {journal} {Nuclear Fusion}\ }\textbf {\bibinfo {volume} {65}},\
  \bibinfo {pages} {052002} (\bibinfo {year} {2025})}\BibitemShut {NoStop}%
\bibitem [{\citenamefont {Smolyakov}\ \emph {et~al.}(2000)\citenamefont
  {Smolyakov}, \citenamefont {Diamond},\ and\ \citenamefont
  {Medvedev}}]{smolyakov2000role}%
  \BibitemOpen
  \bibfield  {author} {\bibinfo {author} {\bibfnamefont {A.}~\bibnamefont
  {Smolyakov}}, \bibinfo {author} {\bibfnamefont {P.}~\bibnamefont {Diamond}},\
  and\ \bibinfo {author} {\bibfnamefont {M.}~\bibnamefont {Medvedev}},\ }\href
  {https://doi.org/10.1063/1.1289514} {\bibfield  {journal} {\bibinfo
  {journal} {Physics of Plasmas}\ }\textbf {\bibinfo {volume} {7}},\ \bibinfo
  {pages} {3987} (\bibinfo {year} {2000})}\BibitemShut {NoStop}%
\bibitem [{\citenamefont {Zholobenko}\ \emph {et~al.}(2021)\citenamefont
  {Zholobenko} \emph {et~al.}}]{zholobenko2021electric}%
  \BibitemOpen
  \bibfield  {author} {\bibinfo {author} {\bibfnamefont {W.}~\bibnamefont
  {Zholobenko}} \emph {et~al.},\ }\href
  {https://doi.org/10.1088/1361-6587/abd97e} {\bibfield  {journal} {\bibinfo
  {journal} {Plasma Physics and Controlled Fusion}\ }\textbf {\bibinfo {volume}
  {63}},\ \bibinfo {pages} {034001} (\bibinfo {year} {2021})}\BibitemShut
  {NoStop}%
\bibitem [{\citenamefont {Diamond}\ \emph {et~al.}(2005)\citenamefont
  {Diamond}, \citenamefont {Itoh}, \citenamefont {Itoh},\ and\ \citenamefont
  {Hahm}}]{diamond2005zonal}%
  \BibitemOpen
  \bibfield  {author} {\bibinfo {author} {\bibfnamefont {P.~H.}\ \bibnamefont
  {Diamond}}, \bibinfo {author} {\bibfnamefont {S.}~\bibnamefont {Itoh}},
  \bibinfo {author} {\bibfnamefont {K.}~\bibnamefont {Itoh}},\ and\ \bibinfo
  {author} {\bibfnamefont {T.}~\bibnamefont {Hahm}},\ }\href
  {https://doi.org/10.1088/0741-3335/47/5/R01} {\bibfield  {journal} {\bibinfo
  {journal} {Plasma Physics and Controlled Fusion}\ }\textbf {\bibinfo {volume}
  {47}},\ \bibinfo {pages} {R35} (\bibinfo {year} {2005})}\BibitemShut
  {NoStop}%
\bibitem [{\citenamefont {Scott}(2005{\natexlab{a}})}]{scott2005energetics}%
  \BibitemOpen
  \bibfield  {author} {\bibinfo {author} {\bibfnamefont {B.~D.}\ \bibnamefont
  {Scott}},\ }\href {https://doi.org/10.1088/1367-2630/7/1/092} {\bibfield
  {journal} {\bibinfo  {journal} {New Journal of Physics}\ }\textbf {\bibinfo
  {volume} {7}},\ \bibinfo {pages} {92} (\bibinfo {year}
  {2005}{\natexlab{a}})}\BibitemShut {NoStop}%
\bibitem [{\citenamefont {Naulin}\ \emph {et~al.}(2005)\citenamefont {Naulin},
  \citenamefont {Kendl}, \citenamefont {Garcia}, \citenamefont {Nielsen},\ and\
  \citenamefont {Rasmussen}}]{naulin2005shear}%
  \BibitemOpen
  \bibfield  {author} {\bibinfo {author} {\bibfnamefont {V.}~\bibnamefont
  {Naulin}}, \bibinfo {author} {\bibfnamefont {A.}~\bibnamefont {Kendl}},
  \bibinfo {author} {\bibfnamefont {O.}~\bibnamefont {Garcia}}, \bibinfo
  {author} {\bibfnamefont {A.}~\bibnamefont {Nielsen}},\ and\ \bibinfo {author}
  {\bibfnamefont {J.~J.}\ \bibnamefont {Rasmussen}},\ }\bibfield  {journal}
  {\bibinfo  {journal} {Physics of Plasmas}\ }\textbf {\bibinfo {volume}
  {12}},\ \href {https://doi.org/10.1063/1.1905603} {10.1063/1.1905603}
  (\bibinfo {year} {2005})\BibitemShut {NoStop}%
\bibitem [{\citenamefont {Grenfell}\ \emph {et~al.}(2024)\citenamefont
  {Grenfell}, \citenamefont {Gil}, \citenamefont {Manz}, \citenamefont {Silva},
  \citenamefont {Adamek}, \citenamefont {Brida}, \citenamefont {Conway},
  \citenamefont {Eich}, \citenamefont {Faitsch}, \citenamefont {Happel} \emph
  {et~al.}}]{grenfell2024multi}%
  \BibitemOpen
  \bibfield  {author} {\bibinfo {author} {\bibfnamefont {G.}~\bibnamefont
  {Grenfell}}, \bibinfo {author} {\bibfnamefont {L.}~\bibnamefont {Gil}},
  \bibinfo {author} {\bibfnamefont {P.}~\bibnamefont {Manz}}, \bibinfo {author}
  {\bibfnamefont {C.}~\bibnamefont {Silva}}, \bibinfo {author} {\bibfnamefont
  {J.}~\bibnamefont {Adamek}}, \bibinfo {author} {\bibfnamefont
  {D.}~\bibnamefont {Brida}}, \bibinfo {author} {\bibfnamefont
  {G.}~\bibnamefont {Conway}}, \bibinfo {author} {\bibfnamefont
  {T.}~\bibnamefont {Eich}}, \bibinfo {author} {\bibfnamefont {M.}~\bibnamefont
  {Faitsch}}, \bibinfo {author} {\bibfnamefont {T.}~\bibnamefont {Happel}},
  \emph {et~al.},\ }\href {https://doi.org/10.1088/1741-4326/ad751b} {\bibfield
   {journal} {\bibinfo  {journal} {Nuclear Fusion}\ }\textbf {\bibinfo {volume}
  {64}},\ \bibinfo {pages} {104002} (\bibinfo {year} {2024})}\BibitemShut
  {NoStop}%
\bibitem [{\citenamefont {Scott}(2021)}]{scott2021turbulence}%
  \BibitemOpen
  \bibfield  {author} {\bibinfo {author} {\bibfnamefont {B.}~\bibnamefont
  {Scott}},\ }\href {https://doi.org/10.1088/978-0-7503-2504-2ch7} {\emph
  {\bibinfo {title} {Turbulence and Instabilities in Magnetised Plasmas, Volume
  1: Fluid drift turbulence}}}\ (\bibinfo  {publisher} {IOP Publishing},\
  \bibinfo {year} {2021})\BibitemShut {NoStop}%
\bibitem [{\citenamefont {Connor}\ \emph {et~al.}(1978)\citenamefont {Connor},
  \citenamefont {Hastie},\ and\ \citenamefont {Taylor}}]{connor1978shear}%
  \BibitemOpen
  \bibfield  {author} {\bibinfo {author} {\bibfnamefont {J.}~\bibnamefont
  {Connor}}, \bibinfo {author} {\bibfnamefont {R.}~\bibnamefont {Hastie}},\
  and\ \bibinfo {author} {\bibfnamefont {J.}~\bibnamefont {Taylor}},\ }\href
  {https://doi.org/10.1103/PhysRevLett.40.396} {\bibfield  {journal} {\bibinfo
  {journal} {Physical Review Letters}\ }\textbf {\bibinfo {volume} {40}},\
  \bibinfo {pages} {396} (\bibinfo {year} {1978})}\BibitemShut {NoStop}%
\bibitem [{\citenamefont {Zeiler}\ \emph {et~al.}(1998)\citenamefont {Zeiler},
  \citenamefont {Biskamp}, \citenamefont {Drake},\ and\ \citenamefont
  {Rogers}}]{Zeiler1998}%
  \BibitemOpen
  \bibfield  {author} {\bibinfo {author} {\bibfnamefont {A.}~\bibnamefont
  {Zeiler}}, \bibinfo {author} {\bibfnamefont {D.}~\bibnamefont {Biskamp}},
  \bibinfo {author} {\bibfnamefont {J.~F.}\ \bibnamefont {Drake}},\ and\
  \bibinfo {author} {\bibfnamefont {B.~N.}\ \bibnamefont {Rogers}},\ }\href
  {https://doi.org/10.1063/1.872953} {\bibfield  {journal} {\bibinfo  {journal}
  {Physics of Plasmas}\ }\textbf {\bibinfo {volume} {5}},\ \bibinfo {pages}
  {2654} (\bibinfo {year} {1998})}\BibitemShut {NoStop}%
\bibitem [{\citenamefont {Scott}(2005{\natexlab{b}})}]{scott2005drift}%
  \BibitemOpen
  \bibfield  {author} {\bibinfo {author} {\bibfnamefont {B.~D.}\ \bibnamefont
  {Scott}},\ }\bibfield  {journal} {\bibinfo  {journal} {Physics of Plasmas}\
  }\textbf {\bibinfo {volume} {12}},\ \href {https://doi.org/10.1063/1.1917866}
  {10.1063/1.1917866} (\bibinfo {year} {2005}{\natexlab{b}})\BibitemShut
  {NoStop}%
\bibitem [{\citenamefont {Eich}\ \emph {et~al.}(2020)\citenamefont {Eich},
  \citenamefont {Manz}, \citenamefont {Goldston}, \citenamefont {Hennequin},
  \citenamefont {David}, \citenamefont {Faitsch}, \citenamefont {Kurzan},
  \citenamefont {Sieglin}, \citenamefont {Wolfrum}, \citenamefont {Team} \emph
  {et~al.}}]{eich2020turbulence}%
  \BibitemOpen
  \bibfield  {author} {\bibinfo {author} {\bibfnamefont {T.}~\bibnamefont
  {Eich}}, \bibinfo {author} {\bibfnamefont {P.}~\bibnamefont {Manz}}, \bibinfo
  {author} {\bibfnamefont {R.}~\bibnamefont {Goldston}}, \bibinfo {author}
  {\bibfnamefont {P.}~\bibnamefont {Hennequin}}, \bibinfo {author}
  {\bibfnamefont {P.}~\bibnamefont {David}}, \bibinfo {author} {\bibfnamefont
  {M.}~\bibnamefont {Faitsch}}, \bibinfo {author} {\bibfnamefont
  {B.}~\bibnamefont {Kurzan}}, \bibinfo {author} {\bibfnamefont
  {B.}~\bibnamefont {Sieglin}}, \bibinfo {author} {\bibfnamefont
  {E.}~\bibnamefont {Wolfrum}}, \bibinfo {author} {\bibfnamefont {A.~U.}\
  \bibnamefont {Team}}, \emph {et~al.},\ }\href
  {https://doi.org/10.1088/1741-4326/ab7a66} {\bibfield  {journal} {\bibinfo
  {journal} {Nuclear Fusion}\ }\textbf {\bibinfo {volume} {60}},\ \bibinfo
  {pages} {056016} (\bibinfo {year} {2020})}\BibitemShut {NoStop}%
\end{thebibliography}
